\documentclass[twocolumn]{aastex62}
\pdfoutput=1 %for arXiv submission
\usepackage{amsmath,amstext}
\usepackage[T1]{fontenc}
\usepackage[utf8]{inputenc}
\usepackage{apjfonts} 
\usepackage[figure,figure*]{hypcap}
\usepackage{amsmath,amssymb,amsfonts,graphicx}
\usepackage{todonotes}
\usepackage{color}
\usepackage{hyperref}
\usepackage{cleveref}
\usepackage{graphicx}
\usepackage{tabularx}
\usepackage{footnote}
\usepackage{subfigure}

\usepackage{booktabs} %jay
\usepackage{multirow} %jay
\usepackage{gensymb} % jay for degree
\usepackage{tablefootnote} %jay

\usepackage{todonotes}   % notes showed

\shorttitle{PHOTOMETRIC AND KINEMATIC STUDY OF SAI 44 AND SAI 45}
\shortauthors{Maurya et al.}

\received{XXXX, 2021}
\submitjournal{AJ}
\begin{document}

   \title{Photometric and Kinematic study of the open clusters  SAI 44 and SAI 45}

\author{Maurya, Jayanand}
\affiliation{Aryabhatta Research Institute of observational sciencES (ARIES). Nainital, Uttrakhand, India}
\affiliation{School of Studies in Physics and Astrophysics, Pandit Ravishankar Shukla University, Chattisgarh 492 010, India}
\email{jayanand@aries.res.in}

\author{Joshi, Y. C.}
\affiliation{Aryabhatta Research Institute of observational sciencES (ARIES). Nainital, Uttrakhand, India}

\author{Elsanhoury, W. H.}
\affiliation{Astronomy Department, National Research Institute of Astronomy and Geophysics (NRIAG), 11421, Helwan, Cairo, Egypt}
\affiliation{Physics Department, Faculty of Science and Arts, Northern Border University, Turaif Branch, Saudi Arabia}
\author{Sharma, Saurabh}
\affiliation{Aryabhatta Research Institute of observational sciencES (ARIES). Nainital, Uttrakhand, India}
\begin{abstract}

   {}
   {We carry out detailed photometric and kinematic study of the poorly studied sparse open clusters SAI 44 and SAI 45 using ground based BVR$_{c}$I$_{c}$ data supplemented by archival data from \textit{Gaia} eDR3 and Pan-STARRS. The stellar membership are determined using a statistical method based on \textit{Gaia} eDR3 kinematic data and we found 204 members in SAI 44 while only 74 members are identified in SAI 45. The average distances to  SAI 44 and SAI 45 are calculated as 3670$\pm$184 and 1668$\pm$47 pc. The logarithmic age of the clusters are determined as 8.82$\pm$0.10 and 9.07$\pm$0.10 years for SAI 44 and SAI 45, respectively. The color-magnitude diagram of SAI 45 hosts an extended main-sequence turn-off (eMSTO) which may be originated through differential rotation rates of member stars. The mass function slopes are obtained as -1.75$\pm$0.72 and -2.58$\pm$3.20 in the mass rages 2.426-0.990 M$_{\odot}$ and 2.167-1.202 M$_{\odot}$ for SAI 44 and SAI 45, respectively. SAI 44 exhibit the signature of mass segregation while we found a weak evidence of the mass segregation in SAI 45 possibly due to tidal striping. The dynamical relaxation times of these clusters indicate that both the clusters are in dynamically relaxed state. Using AD-diagram method, the apex coordinates are found to be (69$\fdg79\pm0\fdg11$, -30$\fdg82\pm0\fdg15$) for SAI 44 and (-56$\fdg22\pm0\fdg13$, -56$\fdg62\pm0\fdg13$) for SAI 45. The average space velocity components of the clusters SAI 44 and SAI 45 are calculated in units of km s$^{-1}$ as (-15.14$\pm$3.90, -19.43$\pm$4.41, -20.85$\pm$4.57) and (28.13$\pm$5.30, -9.78$\pm$3.13, -19.59$\pm$4.43)}, respectively.
 
  
\end{abstract}


\section{Introduction} \label{sec:intro}
Most of the stars, if not all, are born in clusters in our Galaxy \citep{2003ARA&A..41...57L,2010ARA&A..48..431P}. Therefore study of star clusters is important in order to understand star formation and stellar evolution. Open star clusters are relatively younger systems and witness recent star formation events. These clusters are mostly found in the spiral arms of the Milky Way galaxy therefore they are suitable tracers in the studies of the Galactic disc formation and structures \citep{1998MNRAS.296.1045C, 2003AJ....125.1397C, 2016A&A...593A.116J}. Open clusters also provide ideal testing environment for star formation and evolution theories as stars belonging to clusters are born from almost similar initial conditions and have similar age, distance, and chemical composition \citep{2003ARA&A..41...57L}.
 
The intermediate age open clusters in the Milky Way have been considered as clusters hosting single Main Sequence (MS) i.e. coeval populations. However, few recent studies claim presence of the extended Main Sequence Turn Off (eMSTO) caused by spread in the age due to existence of multiple populations in the Galactic open clusters \citep{2019MNRAS.490.2414P, 2019ApJ...887..199G, 2021ApJ...906..133L}. The age spread inferred from eMSTO were thought to be present due to extended star formation in the clusters \citep{2014ApJ...797...35G} but detection of extended star formation responsible for eMSTO have been elusive \citep{2013MNRAS.436.2852B}. There have been other theories like variability, rotation, and binary-systems explaining the origin of eMSTO \citep{2017NatAs...1E..11D,2019ApJ...876...65L,2021MNRAS.502.4350S}. The most discussed reason behind presence of eMSTO is the rotation of stars affecting effective temperature of stars, thus causing spread in the color of stars in upper MS \citep{2019A&A...622A..66G}. The fast rotating stars have been found to be located in the red part of eMSTO while slow rotating stars are located in blue part of eMSTO \citep{2019ApJ...876..113S}. The origin of eMSTO in the Galactic open clusters is currently intensely debated and this makes these clusters even more interesting for investigations. The dynamical evolution of stars in the clusters are important in the study of initial mass distribution and dynamics of the Galactic disc. Mass functions and dynamical evolution of the clusters can be estimated once the physical parameters like age and distance modulus are accurately determined. The various aspects of the initial mass function like spatial variation and universality with time are still open questions to conclude \citep{2010ARA&A..48..339B,2018A&A...614A..43D}. Sometimes spatial variation in mass function slopes indicates mass segregation in open clusters \citep{2019MNRAS.489.2377S}. There are two theories used to explain mass segregation phenomenon: primordial mass segregation and segregation caused by dynamical evolution of clusters. According to primordial theory massive stars are formed preferentially in the central part of the cluster \citep{2008ApJ...678L.105D}. The other theory suggests that masss segregation in the cluster is caused by two-body relaxation \citep{2009MNRAS.395.1449A}. It is still debated which one or both the mechanism are responsible for the segregation of massive stars in clusters \citep{2018MNRAS.473..849D}. Thus comprehensive photometric study of open clusters is useful in understanding the physics of stars from star formation events to the stellar and dynamical evolution of stars.

In this study, we present photometric study of unstudied open clusters SAI 44 and SAI 45. These two clusters are listed in the catalog compiled by research group at Sternberg Astronomical Institute (SAI), Russia \citep{2008A&A...486..771K,2010AstL...36...75G}. The catalog have initial parameters of newly discovered open clusters using 2MASS data \citep{2006AJ....131.1163S}. These two clusters are not photometrically studied in a great detail and our photometric study based on precise membership analysis using kinematic data of \textit{Gaia} eDR3 \citep{2020arXiv201201533G, 2020arXiv201203380L} will enrich the study of these clusters. 

In this paper, data analysis is given in Section~\ref{data}. The spatial distribution of the stellar density and radii of these clusters are determined in Section~\ref{RDP}. The membership determination using \textit{Gaia} eDR3 kinematic data and estimation of physical parameters are described in Section~\ref{cluster_param}. The analysis of eMSTO present in the upper main-sequence of SAI 45 is given in Section~\ref{eMSTO}. The MF and dynamic evolution are discussed in Section~\ref{dynamic}. The kinematic study of the clusters is given in Section~\ref{kinematic}. Section~\ref{conclusion} is dedicated to discussion and conclusion of the study.
%__________________________________________________________________

\section{Data}\label{data}
%------------------------------------------------------------------
We used archived data from \textit{Gaia} eDR3, 2MASS, Pan-STARRS surveys to complement our observed data in BVR$_{c}$I$_{c}$ bands. The kinematic data from \textit{Gaia} eDR3 have been particularly useful in the determination of membership, distance, and kinematic structure parameters. The 2MASS data in near-IR J, H, and K$_{s}$ bands were useful in the calculation of total-to-selective extinction R$_{v}$. We converted 2MASS K$_{s}$ band magnitude to K magnitude using relation given by \citet{2001AJ....121.2851C}.
\subsection{Observed  BVR$_{c}$I$_{c}$ data}
\begin{figure}
\centering
  \includegraphics[width=9.0cm, height=7.0cm]{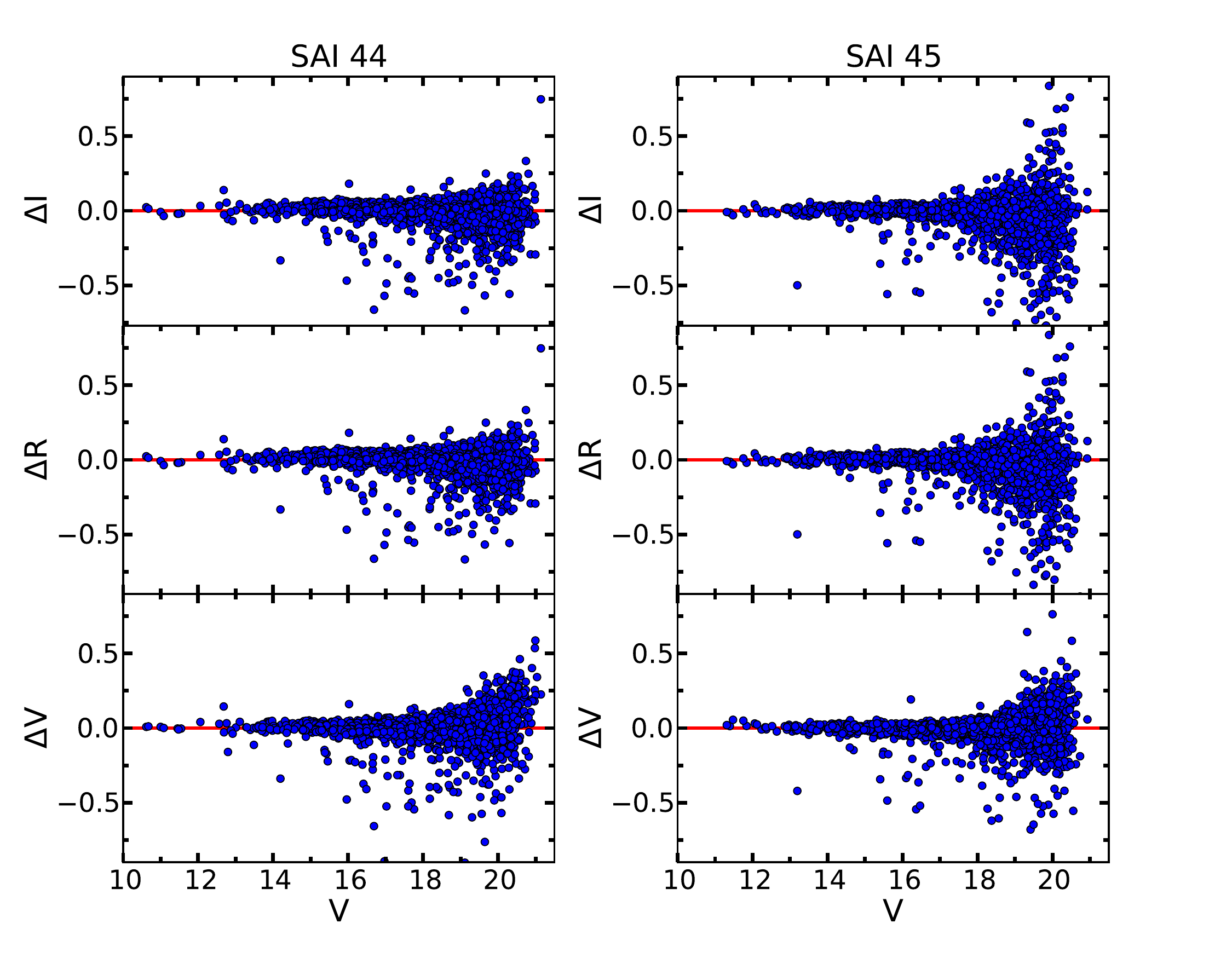} 
  \vspace{-0.6cm}
\caption{Plots of comparison for observed VR$_{c}$I$_{c}$ magnitudes and calculated VR$_{c}$I$_{c}$ magnitudes from \textit{Gaia} DR2 G, G$_{BP}$, and G$_{RP}$ magnitudes using Carrasco conversion formula for SAI 44 and SAI 45.} 
\label{comp_sai}
\end{figure}

The observations in  BVR$_{c}$I$_{c}$ bands of SAI 44 and SAI 45 were captured using 1.3-m Devasthal Fast Optical Telescope (DFOT) situated at an Himalayan site known as Devasthal in India. The observations of SAI 44 were taken on 24 March 2017 while SAI 45 was observed on 25 March 2017. The observations of SAI 44 were taken in VR$_{c}$I$_{c}$ bands and SAI 45 was observed in BVR$_{c}$I$_{c}$ bands. The exposure time for B, V, R$_{c}$, and I$_{c}$ bands were 300, 200, 100, and 60 sec, respectively. The DFOT has a field of view of ~$18^{'}$ $\times$ $18^{'}$ for the 2k$\times$2k CCD camera used for the observations. To standardize instrumental magnitudes and find standard magnitudes, observations of Landolt's standard field SA 98 were taken on the both nights. The observations of astronomical twilight flats and bias images were also acquired on each night. We used Image Reduction and Analysis Facility (IRAF) to clean our images. The instrumental magnitudes were estimated through PSF techniques using DAOPHOT II package. We calibrated the instrumental magnitudes to find the standard magnitudes through process given in \citet{1992ASPC...25..297S}. The transformation equations used for the standardization are given in our previous paper \citep{2020MNRAS.494.4713M}. We calculated VR$_{c}$I$_{c}$ magnitudes from \textit{Gaia} DR2 G, G$_{BP}$, and G$_{RP}$ magnitudes using Carrasco\footnote{\url{https://gea.esac.esa.int}} conversion formula. We have shown comparison between observed VR$_{c}$I$_{c}$ magnitudes and the calculated VR$_{c}$I$_{c}$ magnitudes in Figure~\ref{comp_sai}. We found that the two magnitudes are in good agreement. We also estimated completeness of observed data in V band using method described in \citet{2020MNRAS.494.4713M}. The V band data was found to be complete upto 19 mag for both the clusters SAI 44 and SAI 45.

%---------------------------------------------------------------------
%\input{table01}
%----------------------------------------------------------------

%----------------------------------------------------------------------
%----------------------------------------------------------------------
%
\subsection{\textit{Gaia} eDR3 data}
The \textit{Gaia} eDR3 \citep{2020arXiv201201533G} data are used in this study for membership determination and distance estimation. The \textit{Gaia} eDR3 provides celestial positions and magnitude in G bands up to 21 mag for 1.8 billion sources. It provides parallaxes, proper motions, and the colors (G$_{BP}$-G$_{RP}$) for 1.5 billion of those sources. In \textit{Gaia} eDR3 data, the uncertainties in parallaxes are up to 0.02-0.03 mas for sources with G < 15 while there is uncertainty of 0.07 mas for objects of 17 mag brightness in G band. The uncertainties are significantly higher for fainter stars as uncertainties gets up to 0.5 mas for sources having 20 mag in G band while uncertainty rises  up to 1.3 mas for objects with G=21 mag. The uncertainty in \textit{Gaia} eDR3 proper motion components has also improved to be 0.02-0.03 mas yr$^{-1}$ for G magnitudes less than 15 mag. The uncertainty becomes 0.07 yr$^{-1}$ for 17 mag in G band while 0.5 mas yr$^{-1}$ for objects having G band magnitude equal to 20 mag. The uncertainty becomes 1.4 mas yr$^{-1}$ for objects having 21 mag in G band.
\subsection{Pan-STARRS data}
The Pan-STARRS used 1.8 meter telescope in its first part of survey (PS1) to image the northern sky in g, r, i, z, and y filters. The PS1 survey data have seeings of 1.31, 1.19, and 1.11 arcsec in g, r, and i, respectively \citep{2016arXiv161205242M}. The survey includes faint stars up to $\sim$23.2, 23.2, and 23.1 mag for stacked images in g, r, and i bands, respectively \citep{2016arXiv161205242M}. We calculated Johnson B band magnitudes for stars in SAI 44 from g and r magnitudes using conversion formula given by \citet{2018BlgAJ..28....3K}. \textbf{These} B band magnitudes for SAI 44 were used throughout the present study.
%--------------------------------------------------------------------------------
%
\section{Spatial stellar distribution}\label{RDP}
The study of the spatial structure of open cluster sheds light on the stellar distribution in the cluster according to the mass (or brightness) of stars \citep{2014MNRAS.437..804J}. We calculated radial stellar density distributions in the selected clusters to obtain center and radii of the clusters. We used stars detected in V band to calculate the stellar density. The cluster centers were estimated using maximum density method by assuming cluster center to be the point where stellar density is maximum. Our estimated cluster centers are (05:11:10.51; +45:42:10.25) and (05:16:29.45; +45:35:35.89) for SAI 44 and SAI 45, respectively. \citet{2010AstL...36...75G} calculated the cluster centers using 2MASS data as (05:11:07.4; +45:43:09) and (05:16:35.0; +45:34:56) for SAI 44 and SAI 45, respectively. The right ascension (RA) of the cluster centers are in agreement with those given by \citet{2010AstL...36...75G} while Declination (DEC) of the cluster centers are slightly different from the values reported by \citet{2010AstL...36...75G}. 

\begin{figure}
  %\vspace{-0.4cm}
   \centering
  \includegraphics[width=9.0 cm,height=9.0 cm]{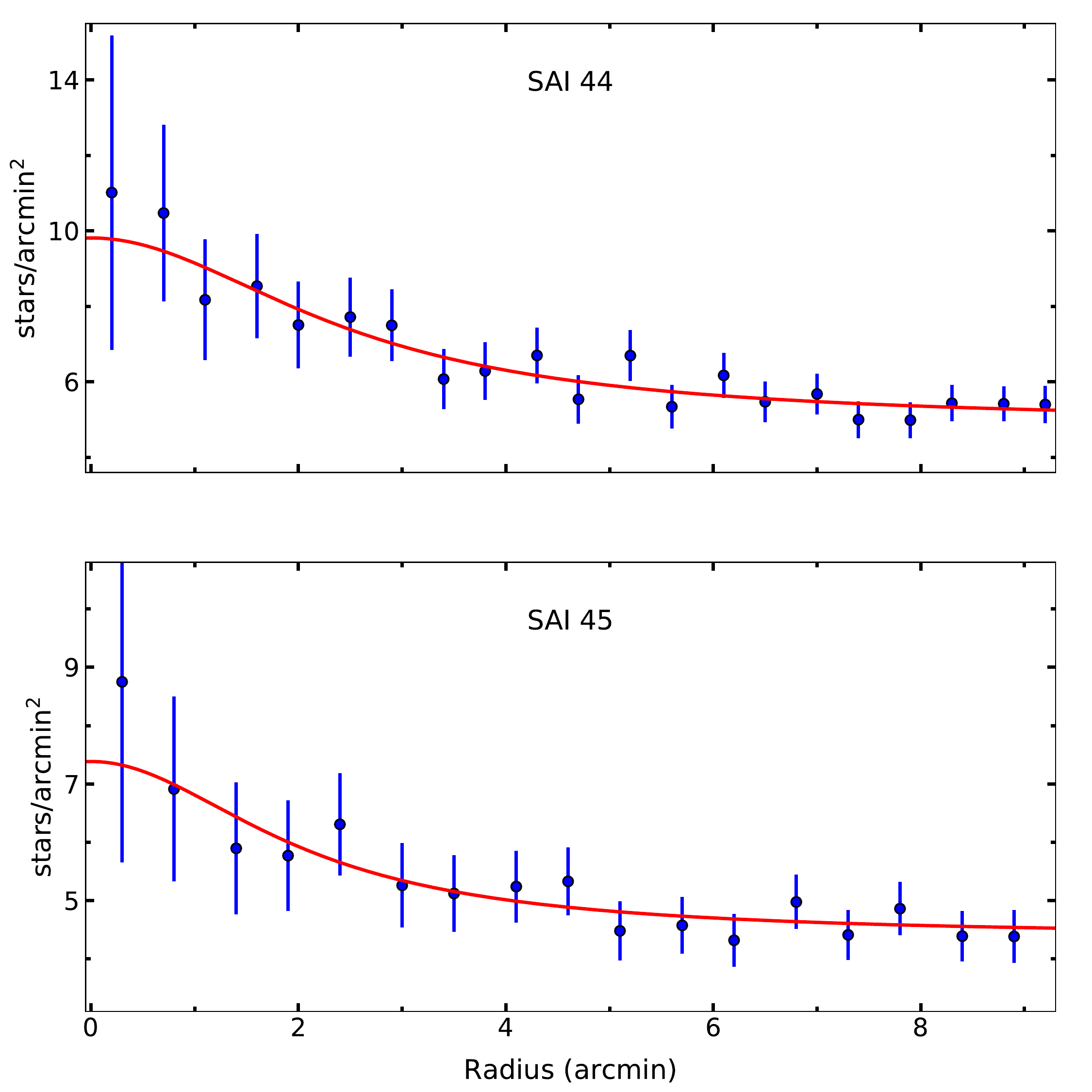}
  \vspace{-0.5cm}
 \caption{The radial density profile for SAI 44 and SAI 45. The best fit red curve presents the empirical radial density profile.}
 \label{rdp}
\end{figure}

The cluster radii were calculated by fitting radial density profile (RDP) provided by \citet{1962AJ.....67..471K} on radial stellar distribution of stars in the observed regions of the selected clusters. We calculated radial stellar distributions in these clusters using concentric annular regions of $\sim$0$^{\prime}$.5 width. The RDP profile given by \citet{1962AJ.....67..471K} is given below: 
$$
\rho(r) = \rho_{b} + \frac{\rho_{0}}{1 + \left(\frac{r}{r_{c}}\right)^{2}}
$$
where $\rho_{0}$ and $\rho_{b}$ are the maximum stellar density and background stellar density, respectively.  r$_{c}$ is the core radius defined by the distance from the center at which stellar density falls to half of its maximum value. The plots of RDPs with fitted \citet{1962AJ.....67..471K} radial density model are shown in Figure~\ref{rdp} for SAI 44 and SAI 45. We found core radii to be 2$^{\prime}$.5$\pm$0$^{\prime}$.6 and 2$^{\prime}$.1$\pm$0$^{\prime}$.8 for SAI 44 and SAI 45, respectively. We considered the cluster boundary to be the points where radial density is 1$\sigma_{b}$ above the background density $\rho_{b}$. The radius of the cluster, r$_{cluster}$, is calculated using the following formula:
$$
r_{cluster} = r_{c} \sqrt{\frac{\rho_{0}}{\sigma_{b}} - 1}
$$
%
%-------------------------------------------------------
%-------------------------------------------------
%
%-------------------------------------------------------
\begin{figure}
   \centering
  \includegraphics[width=9.0 cm,height=5.0 cm]{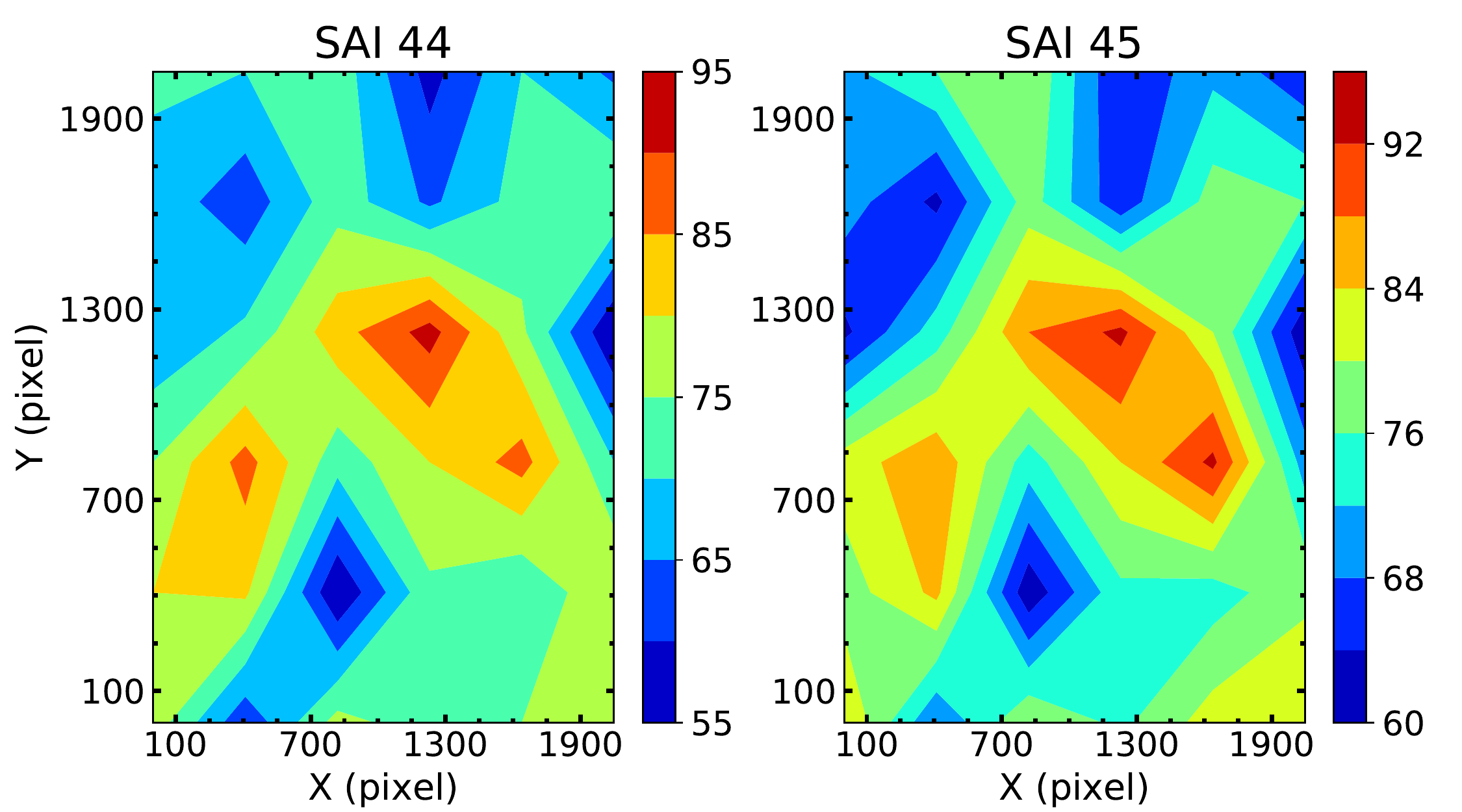}
 \caption{The contour maps for spatial stellar density distribution in observed regions of SAI 44 and SAI 45. The color-bar denotes stellar number density per arcmin$^{2}$ area.}
 \label{contour_rdp}
\end{figure}
%-------------------------------------------------
%

We obtained cluster radii as 11$^{\prime}$.4$\pm$0$^{\prime}$.5, and 8$^{\prime}$.9$\pm$0$^{\prime}$.5 for the clusters SAI 44 and SAI 45, respectively. The estimated radii are approximate in nature due to irregular shape of the clusters as also visible in Figure~\ref{contour_rdp} showing contour maps of spatial density distribution for these clusters. The estimation of cluster radius based on boundary condition depends on the chosen cut off therefore resulting radius is lower limit of possible radius and some stars belonging to the cluster may be found beyond the calculated cluster radius. The stellar density contrast of these clusters against background population can be estimated through density contrast parameter, $\delta_{c}$, which is mathematically defined as follow:
$$
\delta_{c} = 1 + \frac{\rho_{0}}{\rho_{b}}
$$
We calculated $\delta_{c}$ to be 2.0 and 1.7 for SAI 44 and SAI 45, respectively. These clusters are sparse clusters relative to their background population density as $\delta_{c}$ for the two studied clusters are below the range (7 $\leq \delta_{c} \leq$ 23) suggested by \citet{2009MNRAS.397.1915B} for the compact clusters. As the ages of open clusters increase stellar dynamics within the clusters influences the central part of the cluster to become circular while the overall clusters extent becomes larger and the stellar density of the clusters becomes sparser \citep{2004AJ....128.2306C}. The clusters SAI 44 and SAI 45 having ages near $\sim$ 1 Gyr are expected to be sparser which is also confirmed from $\delta_{c}$ obtained for these clusters. All these estimated structural parameters are included in Table~\ref{struc_par} for both the clusters.

%--------------------------------------------------------------------------------
\begin{table}
  \centering
  \caption{The determined structural parameters for the clusters.}
  \label{struc_par}
  \begin{tabular}{c c c c c c}  
  \hline  
  Cluster &  \multicolumn{2}{c}{Central coordinates} & r$_{cluster}$ & $\delta_{c}$\\
          &  RA (J2000)      & DEC (J2000)           &   ($\prime$)  &             \\ \hline
   SAI 44 &05:11:10.51&+45:42:10.25&11$^{\prime}$.4$\pm$0$^{\prime}$.5&2.0\\
   SAI 45 &05:16:29.45&+45:35:35.89&8$^{\prime}$.9$\pm$0$^{\prime}$.5&1.7\\ \hline
  \end{tabular}
\end{table}

%-------------------------------------------------------------------

%--------------------------------------
\section{Member stars and cluster parameters}\label{cluster_param}
%--------------------------------------
%--------------------------------------
\subsection{Membership}\label{membership}
%--------------------------------------
%--------------------------------------------------------------
\begin{figure*}
   \centering
\vspace{-1.0cm}
  \includegraphics[width=15 cm,height=11.0 cm]{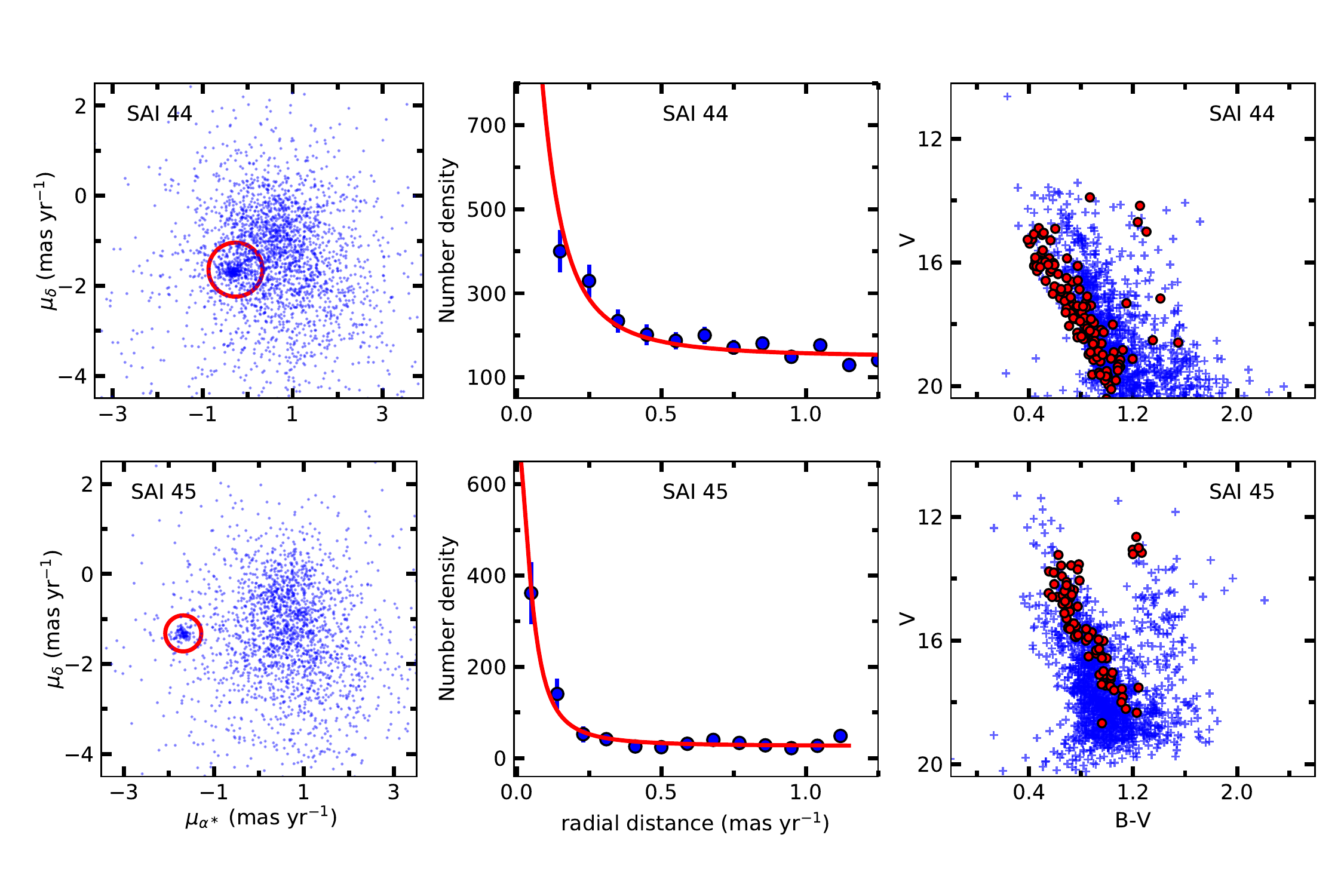}
  \vspace{-0.8cm}
 \caption{The plots of Vector-point diagram (left panel), stellar density distribution as a function of radial distance in proper motion plane (middle panel), and color-magnitude diagrams (right panel) for SAI 44 and SAI 45. In the vector-point diagrams the probable member stars are encircled within red circle. The red circular points and blue '+' marks denote the member stars and field stars, respectively in the color-magnitude diagrams.}
 \label{vpd_rdp}
\end{figure*}
%-------------------------------------------------------------

Membership assessment of stars in a cluster region is of central importance in detailed analysis of the cluster. The stars belonging to a cluster are gravitationally bound together therefore they are expected to have proper motions tightly distributed around the mean proper motion value. Thus, the distribution of stars in proper motion plane have been very usefully in membership determination after advent of precise astrometric data of \textit{Gaia} DR2 and eDR3 \citep{2018A&A...618A..93C,2020MNRAS.499.1874M}. We utilized this property of distribution of stars in proper motion plane through vector-point diagrams (VPDs) as initial step to separate cluster stars from field stars. The VPDs of these clusters are shown in Figure~\ref{vpd_rdp}. It is clear from Figure~\ref{vpd_rdp} that cluster stars are conspicuously separated from field stars in VPDs of the selected open clusters. The stars lying in the red circle in the VPDs are probable cluster members. We derived the center of red circle using maximum density method for the stars in proper motion plane. We also calculated radial distribution of number density of probable member stars in proper motion plane which is shown in Figure~\ref{vpd_rdp}. The fitted red curves on radial density distributions of probable members in proper motion plane are similar to the curve fitted on RDP in spatial plane in Sect.~\ref{RDP}. The radii of the red circles in VPDs of the clusters are also estimated from the radial density distribution in proper motion plane. The radii were taken as radial distance up to the point where cluster stellar density starts merging with field star density. We found centers of the circles in VPDs to be (-0.21, -1.62) and (-1.68, -1.33) in mas yr$^{-1}$ units for SAI 44 and SAI 45, respectively. The radii of the circles were chosen to be 0.6 and 0.4 mas yr$^{-1}$ for SAI 44 and SAI 45, respectively. There are 420 and 101 probable member stars encircled in VPDs of the clusters SAI 44 and SAI 45, respectively.

\begin{figure}
  \includegraphics[width=8.5 cm,height=5.0 cm]{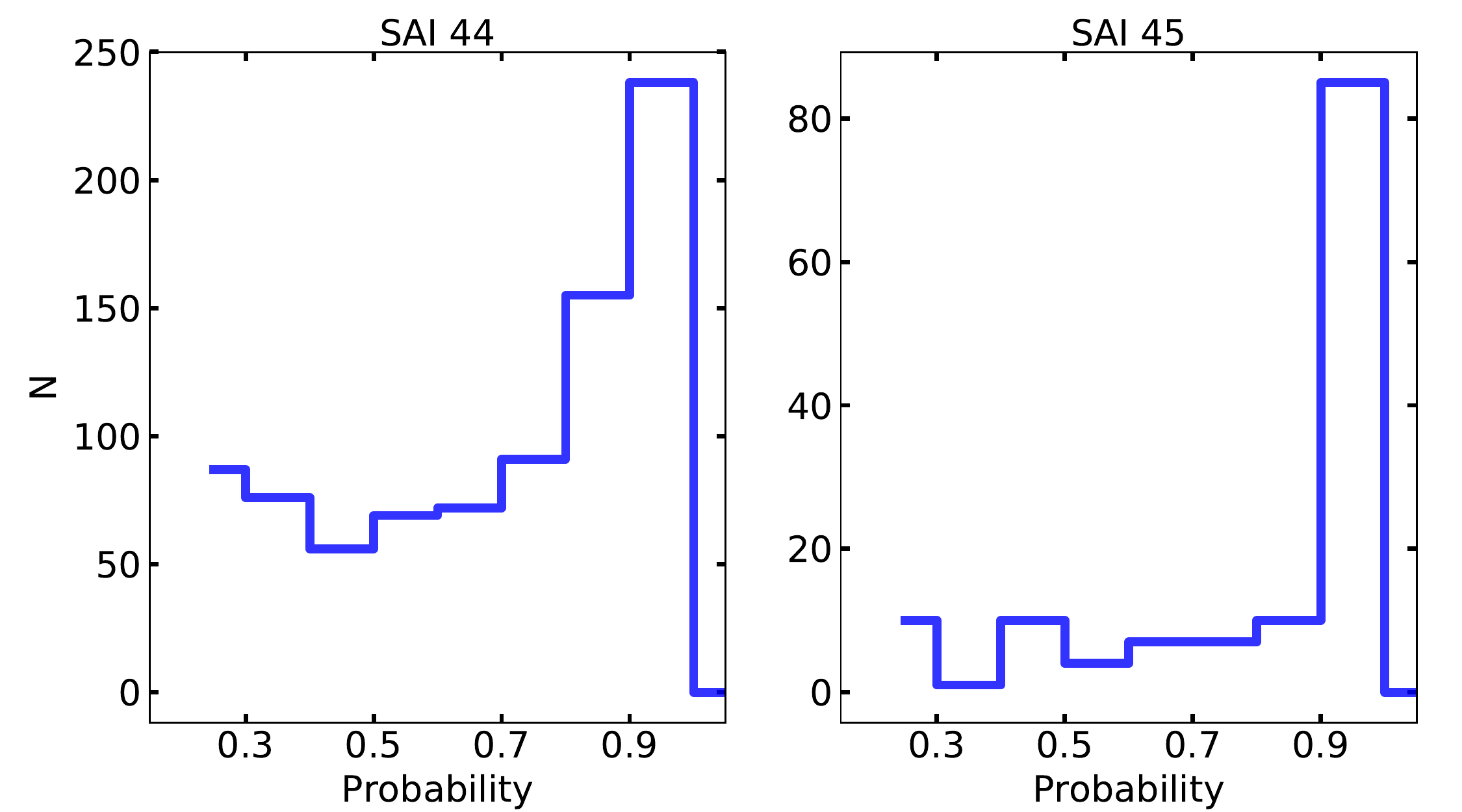}
 \caption{Histograms of probability distribution in SAI 44 and SAI 45.}
 \label{pb_cuts}
\end{figure}

We calculated membership probabilities for the stars present in the observed regions of the selected clusters through statistical method using PMs of stars as given in previous studies \citep{1971A&A....14..226S,2020MNRAS.492.3602J}. The equations used in membership calculation using this method are given in our previous study \citep{2020MNRAS.494.4713M}. The region belonging to the cluster stars in VPD for SAI 45 is separated very clearly from field stars region while separation of the two regions is not quite well for SAI 44. Thus, probability cut-off for the two clusters are expected to be different as the method of membership probability calculation is based on the distribution of stars in the proper motion plane \citep{1971A&A....14..226S}. The cluster stars are known to have very similar proper motions therefore they would be concentrated in only few bins in the histogram of probability distribution. We plotted histograms of probability distribution as shown in Figure~\ref{pb_cuts}. The number of stars started showing clear upward trend from 70$\%$ and 90$\%$ for SAI 44 and SAI 45, respectively. Therefore, we took probability cut-off for SAI 44 and SAI 45 to be 70$\%$ and 90$\%$, respectively. However, there is still possibility that stars satisfying probability cut-off may belongs to field population as few field stars may have proper motion similar to the cluster stars. The most reliable way to remove these field stars from the cluster stars sample is the application of parallax cut-off because stars belonging to the cluster are expected to have very tight parallax distribution \citep{2020MNRAS.495.2496M}. We took parallax cut-off within 1$\sigma$ of mean parallax value ($\bar{\varpi}$) calculated from probable members for SAI 44 while within 3$\sigma$ of the mean parallax was used as selection criteria for SAI 45. The cluster SAI 44 have smaller $\bar{\varpi}$ and larger $\sigma$ than SAI 45 so we applied narrow selection criteria in parallax for SAI 44 as compared to SAI 45. We identified 204 and 74 member stars in SAI 44, and SAI 45, respectively after applying both the probability cut-off and the parallax cut-off.

%--------------------------------------------------------------
%-------------------------------------------------------------

In order to evaluate the cluster parameters with high confidence level, all the further analysis in this study have been carried out considering only these member stars unless explicitly mentioned otherwise. The average PM values ($\bar{\mu}_{\alpha*}$, $\bar{\mu}_{\delta}$) are calculated to be (-0.20$\pm$0.25, -1.63$\pm$0.22), and (-1.68$\pm$0.08, -1.33$\pm$0.08) in mas yr$^{-1}$ units for  SAI 44 and SAI 45, respectively. \citet{2020A&A...640A...1C} found 80 and 27 member stars having magnitude only upto G=17 mag in SAI 44 and SAI 45, respectively. The mean PM values ($\bar{\mu}_{\alpha*}$, $\bar{\mu}_{\delta}$) of (-0.265; -1.628) and (-1.609; -1.288) for  SAI 44 and SAI 45, respectively calculated by \citet{2020A&A...640A...1C} are in good agreement with our estimated values of mean PMs.

\subsection{PHYSICAL PARAMETERS}\label{param}
The accurate determination of physical parameters of the clusters plays important role in the study of stellar and dynamical evolution in the clusters as well as evolution and structure of the Galaxy disc \citep{2008AJ....136..118F}. The determination of the distance of a cluster is important to know its location which is in turn related to properties like cluster morphology and reddening \citep{2019ApJ...887...93G, 2021arXiv210302912H}. The distance calculated through parallax are independent of intrinsic properties of the objects unlike in the case of isochrone fitting. The accuracy of kinematic data including parallax has improved very significantly in \textit{Gaia} eDR3 data in comparison to \textit{Gaia} DR2 kinematic data \citep{2020arXiv201203380L} so we used \textit{Gaia} eDR3 parallax to determine distance to the clusters SAI 44 and SAI 45. The distances to these clusters were calculated using the method suggested by \citet{2018AJ....156...58B}\footnote{\url{https://github.com/ehalley/Gaia-DR2-distances}} which uses a Bayesian approach using a prior of exponentially decreasing space density. We applied systematic offset of -0.017 mas reported by \citet{2020arXiv201201742L} to find offset corrected mean parallax as 0.272 and 0.600 mas for SAI 44, and SAI 45, respectively. The distances to the clusters  SAI 44 and SAI 45 are found to be 3670$\pm$184 and 1668$\pm$47 pc, respectively. \citet{2020A&A...640A...1C} reported mean parallax and distance as 0.252 mas and 3575 pc respectively for SAI 44. The mean parallax and distance of the cluster SAI 45 is given to be 0.561 mas and 1694 pc, respectively in \citep{2018A&A...618A..93C}.

Although the value of total to selective extinction R$_{v}$ is taken to be 3.1 for diffused interstellar medium \citep{1989ApJ...345..245C}, the value of R$_{v}$ have been found to vary in the Galaxy for different line of sights \citep{2004ApJ...616..912V}. The total to selective extinction is very helpful in photometric study of clusters as it allows the calculation of extinction A$_{v}$ directly from the reddening E(B-V) which is easily measured parameter of the clusters. We obtained R$_{v}$ as 3.1 and 2.8 for SAI 44 and SAI 45, respectively from the slopes of color-color diagrams using optical and near-IR J, H, and K bands data. The method to calculate R$_{v}$ is described in \citet{2020MNRAS.494.4713M}. The R$_{v}$ values have been used to infer the size of dust grains \citep{2016ApJ...821...78S}. The lower value of total-to-selective extinction for SAI 45 may indicates smaller dust grain size for the cluster region than normal dust grain size for diffused interstellar medium. \citet{2017ApJ...838...36S} found a correlation between R$_{v}$ values and distances which states nearby dusts are associated to lower R$_{v}$ values than distant dusts. Although SAI 44 and SAI 45 are situated in the approximately same part of sky, the lower R$_{v}$ for SAI 45 is consistent with \citet{2017ApJ...838...36S} as SAI 45 is situated at significantly smaller distance compared to SAI 44. We calculated reddening E(B-V) to be 0.34$^{+0.01}_{-0.07}$ and 0.34$^{+0.03}_{-0.02}$ mag for SAI 44 and SAI 45, respectively utilizing 3D reddening map provided by \citet{2019ApJ...887...93G} which gives reddening map with very good resolution to the north of -30$\degree$ declination using \textit{Gaia} parallax and multiband photometric data from Pan-STARRS and 2MASS. The E(B-V) values were obtained using RA, DEC, distance, and R$_{v}$ values found in this study through the extinction ratios relations given by \citet{2019ApJ...877..116W}. The extinction A$_{v}$ for these clusters were calculated using A$_{v}$ = R$_{v}$ $\times$ E(B-V). We found the values as 1.05 and 0.95 in the direction of the line of sight of the clusters SAI 44 and SAI 45, respectively. The A$_{v}$ value of calculated by us is larger than A$_{v}$ value of 0.86 for SAI 44 given by \citet{2020A&A...640A...1C}. However, our estimated A$_{v}$ value for SAI 44 is in good agreement with A$_{v}$ = 1.065 obtained by \citet{2019A&A...623A.108B} for SAI 44.

\begin{figure*}
\centering
\includegraphics[width=15cm, height=5.0cm]{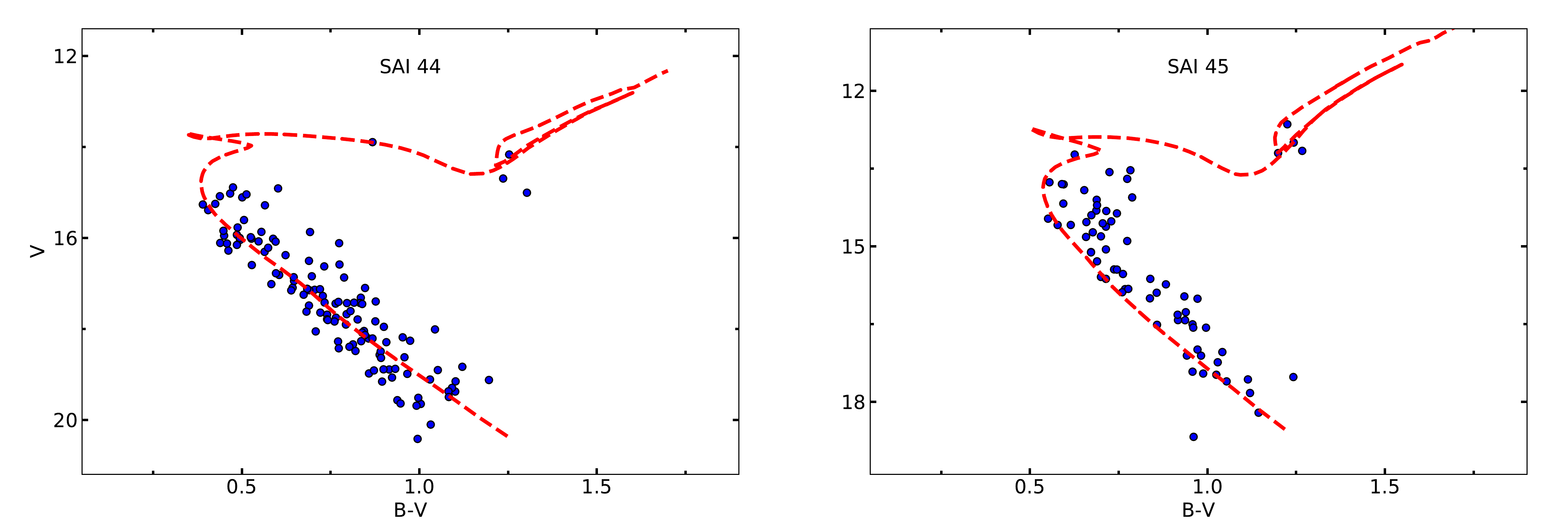}
\includegraphics[width=15cm, height=5.0cm]{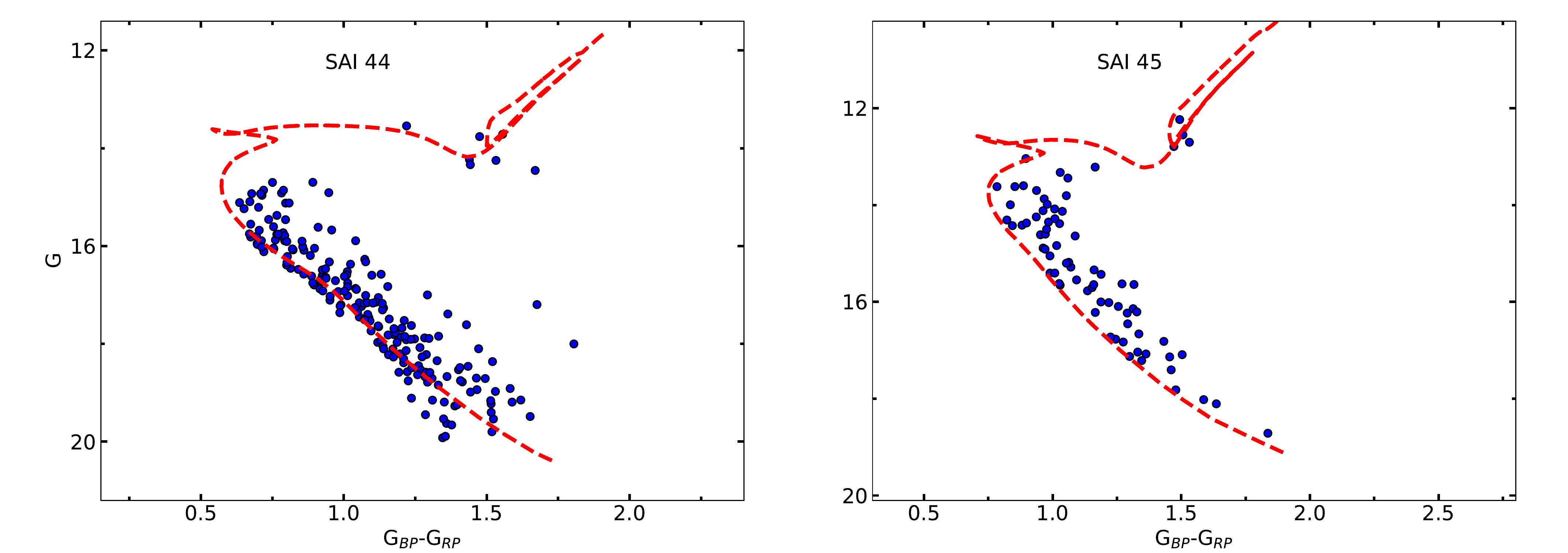}
\caption{Plots of V/(B-V) and G/(G$_{BP}$-G$_{RP}$) color-magnitude diagrams with best fit \citet{2017ApJ...835...77M} solar metallicity isochrones as shown by red solid curves for SAI 44 and SAI 45.}
\label{bv_v}
\end{figure*}

We used A$_{v}$ values calculated in the present study to find apparent distance modulus and extinction corrected isochrone of \citet{2017ApJ...835...77M}. We used V/(B-V) and G/(G$_{BP}$-G$_{RP}$) color-magnitude diagrams (CMDs) of the clusters SAI 44 and SAI 45 to estimate ages of the clusters. The plots of the CMDs for SAI 44 and SAI 45 are shown in Figure~\ref{bv_v}. We fitted \citet{2017ApJ...835...77M} extinction corrected isochrones of solar metallicity on the CMDs of these clusters for our estimated extinction A$_{v}$. The best fits were achieved for isochrones having logarithmic-scaled age of 8.82$\pm$0.10 and 9.07$\pm$0.10 year for distance modulus 12.82$\pm$0.11 and 11.11$\pm$0.06 mag for the clusters SAI 44 and SAI 45, respectively. However, the isochrone fitted on V/(B-V) CMD seems to be slightly redward shifted which might be introduced during conversion of g and r band magnitudes from Pan-STARRS to the B band magnitude. The log(Age) values reported by \citet{2010AstL...36...75G} are 8.95 and 9.20 years for SAI 44 and SAI 45, respectively. The ages estimated by us are slightly younger than the same found by \citet{2010AstL...36...75G} for the clusters SAI 44 and SAI 45. However, the age obtained in the present study for SAI 44 is in agreement with the age given by \citet{2019A&A...623A.108B}.  
%
%--------------------------------------------------------
%-------------------------------------------------
%------------------------------------------------------------
\subsection{Half mass radius, tidal radius, and cluster structure}\label{tidal}
%------------------------------------------------------------
The half-mass radius is defined as the radial distance containing the half of the total mass belonging to a cluster. The masses of stars for R$_{h}$ calculation were found using \citet{2017ApJ...835...77M} isochrones. We found half mass as 132.935 and 49.530 M$_{\odot}$ while corresponding R$_{h}$ values were obtained to be 4$^{\prime}$.1 (4.4 pc) and 4$^{\prime}$.5 (2.2 pc) for SAI 44 and SAI 45, respectively. The R$_{h}$ value combined with tidal radius, R$_{t}$, is very useful in study of dynamics like extent of cluster disruption due to tidal forces \citep{2021MNRAS.500.4338A}. Open clusters structure and dynamical evolution are influenced by tidal interactions \citep{2010MNRAS.402.1841C}. Open clusters lose stars continuously due to tidal interactions. The tidal radius is calculated using the relation given by \citet{2000ApJ...545..301K} as following:
$$
R_{t} = \left(\frac{M_{C}}{2 M_{G}}\right)^{1/3}\times R_{gc}
$$
In the above equation M$_{C}$ is the total mass of cluster and R$_{gc}$ is the distance to the cluster from the Galactic center. $R_{gc}$ is calculated in Section~\ref{kinematic} as 11.742 and 9.802 kpc for SAI 44 and SAI 45, respectively. M$_{G}$ denotes the mass of the Galaxy contained within R$_{gc}$. R$_{t}$ is the radial distance from the center of a cluster where gravitational field of the cluster is equal to the tidal field of the Galaxy. The value of M$_{G}$ is calculated using \citet{1987ARA&A..25..377G} relation given as:
$$
M_{G} = 2 \times 10^{8} M_{\odot} \left(\frac{R_{gc}}{30pc }\right)^{1.2}
$$
Using the above two equations, we determined tidal radii of SAI 44 and SAI 45 as 10.1 and 6.5 pc respectively. We obtained R$_{h}$/R$_{t}$ ratios to be 0.4 and 0.3 for SAI 44 and SAI 45, respectively. SAI 45 has smaller value of R$_{h}$/R$_{t}$ ratio implying that it has more compact structure and survival possibility \citep{2021MNRAS.500.4338A}. The clusters located at the larger Galactocentric distances are subject to weaker tidal field, and hence these clusters show larger R$_{h}$/R$_{t}$ ratio and have less compact structures. Since SAI 44 is at large R$_{gc}$ value it is expected to be less compact and have larger core radius compared to SAI 45 which is found to be true as it has larger core radius (see, Section~\ref{RDP}). It has been found that core radius and dynamical time ratio of a cluster are negatively correlated \citep{2021MNRAS.500.4338A}. The dynamical time ratio is calculated as $\tau_{dyn}$ = age/t$_{cross}$ where crossing time, t$_{cross}$, is the time scale required for a star belonging to the cluster to complete an orbit across the system. t$_{cross}$ was calculated using formula t$_{cross}$ = R$_{h}$/$\sigma_{v}$ where R$_{h}$ is the half mass radius and $\sigma_{v}$ is velocity dispersion. We obtained velocity dispersion from the space velocities calculated in Section~\ref{kinematic} to be  8.47 and 1.39 km s$^{-1}$ for SAI 44 and SAI 45, respectively. The values of t$_{cross}$ were found to be 0.52 and 1.58 Myr for SAI 44 and SAI 45, respectively. $\tau_{dyn}$ values were found to be 1270 and 743 for SAI 44 and SAI 45, respectively. The obtained values of core radius 2.7 and 1.0 pc for SAI 44 and SAI 45 are comparable to values shown in the core radius-log($\tau_{dyn}$) correlation plot of \citet{2021MNRAS.500.4338A}. Therefore we can conclude that the structure and dynamics of a cluster is influenced by the tidal field interactions.
%
%-------------------------------------------
\section{Presence of \lowercase{e}MSTO in CMD}\label{eMSTO}
%
%---------------------------------------------------------------------
\begin{figure*}
\centering
\includegraphics[width=18 cm, height=6.0cm]{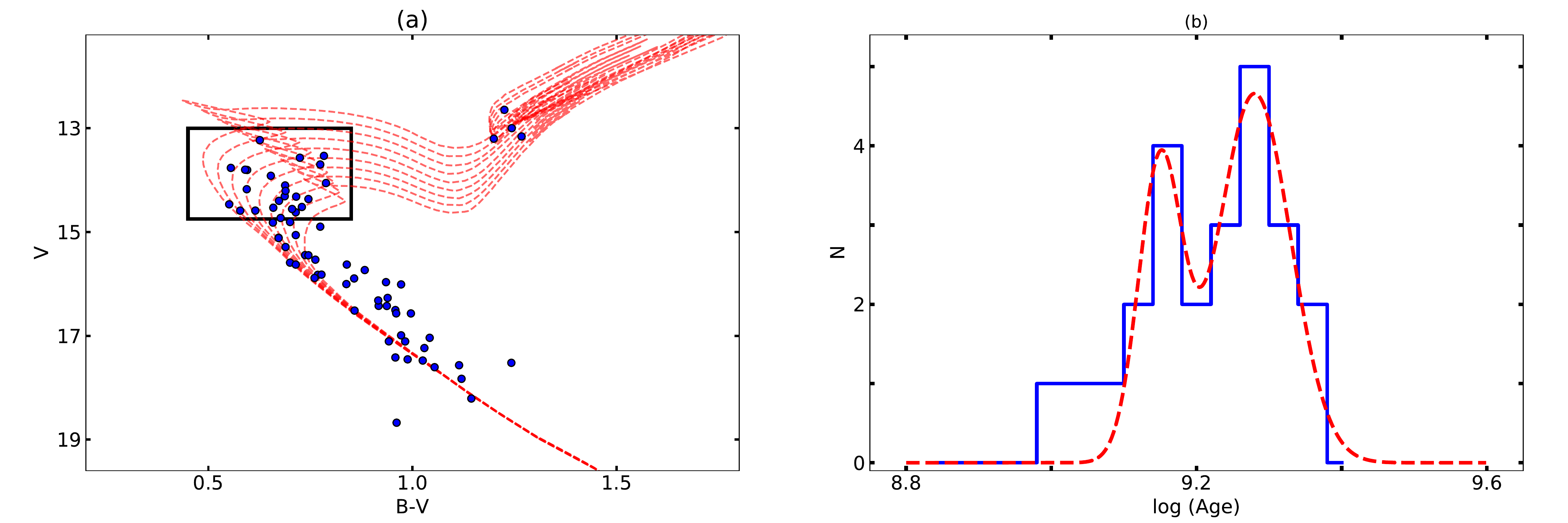}
\caption{(a) color-magnitude diagram for SAI 45 with a rectangular region enclosing extended Main-Sequence Turn-Off. The fitted extinction corrected \citet{2017ApJ...835...77M} isochrones of solar metallicity are shown by red dashed curves. The minimum log(Age) is 9.00 year and the maximum log(Age) is 9.40 year. The log(Age) interval between isochrones in the plots is 0.05 year. (b) Plot of histogram of log(Age) with bin size of log(Age)=0.04 year.}
\label{emsto}
\end{figure*}
%--------------------------------------------------------------------

\begin{figure}
\centering
\includegraphics[width=8.5cm, height=5.0cm]{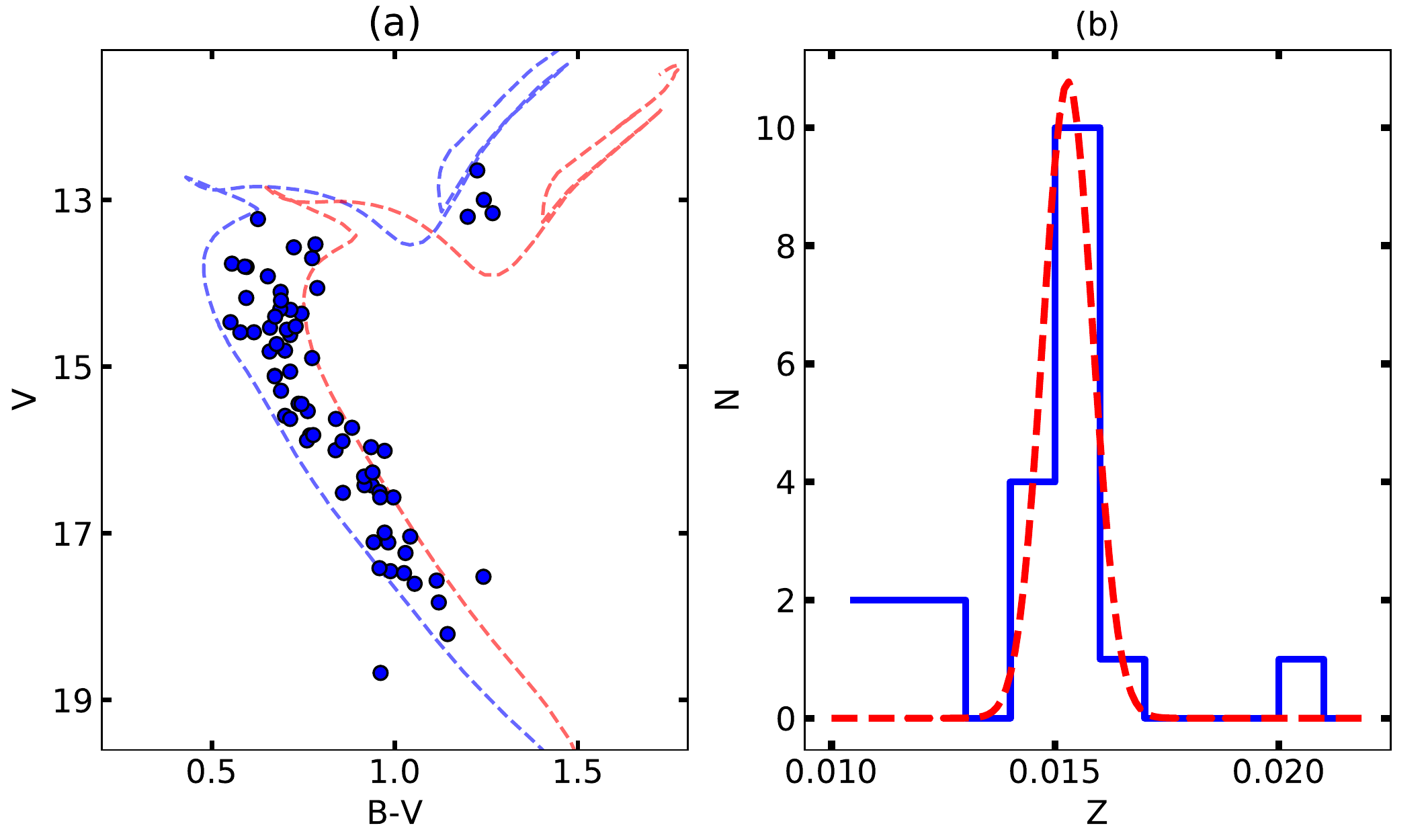}
\caption{(a) color-magnitude diagram fitted with two \citet{2017ApJ...835...77M} isochrones of different metallicity Z. The blue end isochrone is for Z=0.010 and red end isochrone corresponds to Z=0.055. (b) Histogram of metallicity Z with a Gaussian fit shown by red dashed curve.}
\label{hist_met}
\end{figure}

\begin{figure*}
\centering
\includegraphics[width=17cm, height=6.0cm]{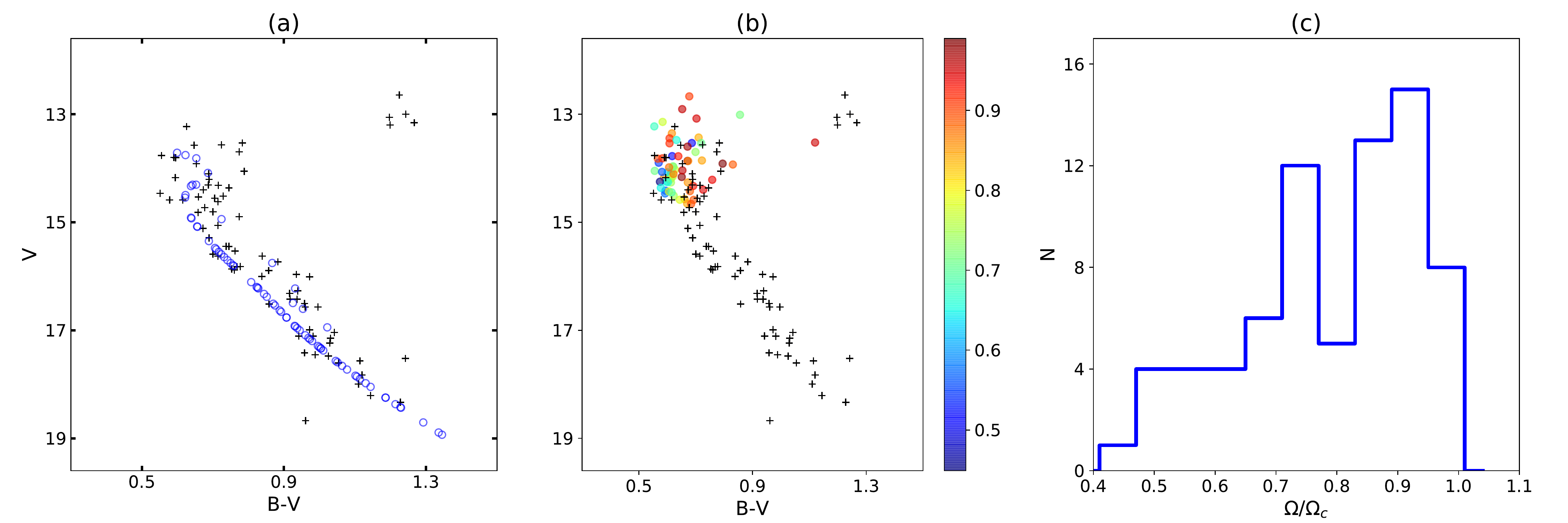}
\caption{(a) color-magnitude diagram with synthetic cluster population over-plotted on observed population. The black '+' signs denote observed population while blue open circles present non-rotating synthetic population generated using SYCLIST interface \citep{2012A&A...537A.146E,2014A&A...566A..21G}. (b) Same as the panel (a) with the only difference that synthetic population includes various rotation rates. The synthetic rotating stars were generated for mass above 1.7 M$_{\odot}$ only. The synthetic population is denoted by filled circle of various colors while '+' sign present observed member stars. The color-bar denotes the values $\Omega/\Omega_{c}$. (c) The histogram plotted using initial rotation rates $\Omega/\Omega_{c}$ of synthetic population.}
\label{emsto_rot}
\end{figure*}

The eMSTO refers to the wider upper MS than usually expected from the single stellar population in a cluster. The cluster SAI 45 shows a broad MS at the brighter end in both V/(B-V) and G/(G$_{BP}$-G$_{RP}$) CMDs which suggests a presence of eMSTO in this cluster. The presence of eMSTO in the Galactic open clusters is relatively recent phenomenon and have been reported in recent studies of open clusters \citep{2018ApJ...869..139C, 2020MNRAS.494.4713M, 2021ApJ...906..133L}. We fitted \citet{2017ApJ...835...77M} isochrone of various ages in Figure~\ref{emsto} but the same metallicity, E(B-V), and distance modulus as used in Figure~\ref{bv_v}. The isochrones of different ages are almost linear in lower part of the main-sequence while they are extended in upper part of the main-sequence. The rectangular region illustrated in Figure~\ref{emsto}(a) is the extended main-sequence region used for calculation of the apparent age spread through isochrone fitting. We have drawn a histogram of apparent age spread for the stars within the rectangular region of the CMD. The plot of the histogram is shown in Figure~\ref{emsto}(b) where a bimodal distribution is clearly evident suggesting an age spread as has also been reported in some previous studies \citep[e.g,][]{2017ApJ...846...22G,2019MNRAS.490.2414P}. The mean logarithmic ages corresponding to the peaks are 9.15 and 9.28 years for small and large peaks, respectively. The apparent age separation between these peaks is 493 Myr. The photometric uncertainty in the magnitudes can also introduce a very small apparent age spread as bright stars belonging to the eMSTO have relatively smaller uncertainty in magnitudes than the fainter stars in the lower MS. We calculated the apparent age spread caused by photometric uncertainty in (B-V) color through the ratio of spread in the age to the spread in (B-V) color for eMSTO stars. The minimum and the maximum apparent logarithmic ages of stars in the eMSTO region are 9.00 and 9.35 years respectively. The total spread in (B-V) color and log(Age) are 0.237 mag and 0.35 year, respectively. The average photometric uncertainty in the (B-V) values is found to be 0.006 mag which can produce an apparent age spread of only $\sim$34 Myr. The spread in upper MS can also be caused by spread in metallicity of stars. We have fitted isochrones with different Z values while keeping other parameters constant as shown in Figure~\ref{hist_met}. The left most isochrone corresponds to Z=0.01 while right most isochrone have Z=0.055. We obtained the [Fe/H] values from \citet{2019A&A...628A..94A} and converted it into Z abundance value using the relation log(Z) = 0.977$\times$[Fe/H] - 1.699 given by \citet{1994A&AS..106..275B}. The histogram of metallicity of stars in eMSTO region with a Gaussian fit is given in Figure~\ref{hist_met}. We obtained the mean and the standard deviation of metallicity Z of eMSTO stars to be 0.0152 and 0.001 respectively. The corresponding FWHM was calculated to be 0.002. The apparent age spread due to spread in metallicity of eMSTO stars was calculated from the ratio of spread in Z require for production of eMSTO to the FWHM of actual distribution of metallicity Z of eMSTO stars. We obtained the apparent spread possible due to metallicity spread to be 22 Myr. Thus, this large apparent spread of 493 Myr has been caused by reasons other than photometric uncertainty in magnitudes and spread in metallicity of eMSTO stars.
%---------------------------------------------------------------------
%---------------------------------------------------------------------
The existance of eMSTO in CMD of open clusters primarily seems to suggest prolonged star formation in the clusters. However, \citet{2015MNRAS.453.2070N} found that the width of eMSTO increases with age of the clusters which suggest eMSTO is related to stellar evolution instead of stellar formation. After subtracting apparent age spread due photometric uncertainty and metallicity spread, the apparent age spread possibly caused by rotation would be 438 Myr. The apparent age spread of 438 Myr is comparable to the previous 300-700 Myr apparent age spread prediction due to rotation for a cluster with age similar to SAI 45 \citep{2015MNRAS.453.2070N,2017ApJ...846...22G}. To further investigate the reason behind the eMSTO, we generated synthetic cluster population using SYCLIST\footnote{\url{https://www.unige.ch/sciences/astro/evolution/en/database/syclist/}} interface \citep{2012A&A...537A.146E,2014A&A...566A..21G}. The synthetic data is over-plotted on the observed CMD in Figure~\ref{emsto_rot}. In Figure~\ref{emsto_rot}(a) the non-rotational synthetic CMD is over-plotted over observed CMD and it fairly reproduces the spread in lower part of main-sequence possibly caused by binaries. However, the  non-rotational synthetic CMD is not able to reproduce the eMSTO. We then used SYCLIST interface to retrieve the main-sequence turn-off only with large number of rotations and over-plotted this synthetic CMD over observed CMD as shown in Figure~\ref{emsto_rot}(b). The eMSTO is nicely reproduced by the synthetic CMD with rotating populations. The fast-rotating stars are preferentially located in red part of the CMD while slow-rotating stars are in the blue part. The gravity darkening in the fast rotating stars causes apparently lower effective temperature \citep{2019ApJ...876...65L} while rotational mixing in the interior of stars causes higher luminosity and cooler temperature \citep{2009MNRAS.398L..11B}. This makes fast rotating stars to appear redder on CMD. There are direct spectroscopic studies too which found that generally fast rotating stars are redder than slow rotating or non-rotating stars \citep{2017ApJ...846L...1D,2018ApJ...863L..33M,2021MNRAS.502.4350S}. We also plotted histogram using initial rotation rates $\Omega/\Omega_{c}$ of the synthetic population as shown in Figure~\ref{emsto_rot}(c). The critical rotation rate $\Omega_{c}$ is the rotation rate at which centrifugal force becomes equal to the surface gravity of stars. The histogram of initial rotation rates exhibits bimodal distribution as obtained from the histogram of apparent age spread shown in Figure~\ref{emsto}(b) . This suggests that the bimodal apparent age distribution of stars in the cluster is mainly caused by differential rotation rates of stars. The bimodal distribution of rotation rates is suggested to be due to slowing down of fast rotating stars caused by tidal braking and binary interaction  \citep{2017NatAs...1E.186D,2021MNRAS.502.4350S}. The synthetic data was generated for solar metallicity Z=0.014, log (Age) = 9.1 year, rotation distribution according to \citet{2010ApJ...722..605H}, and random rotation axis distribution. The limb darkening \citep{2000A&A...359..289C} and gravity darkening \citep{2011A&A...533A..43E} were also accounted in the synthetic cluster population. The binary fraction was chosen to be 0.3 for the synthetic cluster. As stellar rotation mimics the extended region of main-sequence turn-off, eMSTO seems to be mainly stellar evolution phenomenon in SAI 45. The fast rotating stars are expected to be more centrally concentrated \citep{2009MNRAS.398L..11B}. We also noticed that the stars in the red part of the CMD in the eMSTO region i.e. fast rotating stars are mostly inner region stars while the blue part of the CMD i.e. slow rotating stars are dominantly outer region stars (refer to Figure~\ref{m_seg}). This again indicates that spread of eMSTO region is caused mainly by differential rotation of stars in SAI 45. We conclude that the stellar rotations are the main reason behind presence of eMSTO in the cluster SAI 45.
%---------------------------------------------------------------------
%---------------------------------------------------------------------
%-------------------------------------------------------------------
\begin{table}
  \centering
  \caption{The determined luminosity function and mass function for SAI 44 and SAI 45.}
  \label{lf}
  \begin{tabular}{c c c c c c c c c c c}
%   \hspace{-3}  
  \hline
   \hspace{0.3 cm} V \hspace{0.3 cm}& \multicolumn{3}{c}{SAI 44} \hspace{0.3 cm}&\multicolumn{3}{c}{SAI 45} \\
  \cmidrule(lr){2-4}\cmidrule(lr){5-7}
   Range& \hspace{0.3 cm} Mass Range& \hspace{0.3 cm} $\bar{m}$&  N&  Mass Range&  $\bar{m}$&  N \\
    %\hline
 (mag)&  (M$_{\odot}$) &  (M$_{\odot}$) & &  (M$_{\odot}$) &  (M$_{\odot}$) \\
    \hline  
    12-13&            -&      -&   -&  2.167-1.998&  2.107&   2\\
    13-14&  2.459-2.426&  2.452&   1&  1.998-1.711&  1.853&  10\\
    14-15&  2.426-2.086&  2.278&   7&  1.711-1.432&  1.576&  19\\
    15-16&  2.086-1.716&  1.910&  29&  1.432-1.202&  1.298&  16\\
    16-17&  1.716-1.415&  1.568&  44&  1.202-1.015&  1.125&  12\\
    17-18&  1.415-1.149&  1.300&  49&  1.015-0.864&  0.955&  11\\
    18-19&  1.149-0.990&  1.080&  45&  0.864-0.748&  0.818&   3\\   
 \hline
  \end{tabular}
\end{table}

%-------------------------------------------------------------------
\section{Dynamical evolution}\label{dynamic}
% 
%-----------------------------------------------------------
\begin{figure*}
\centering
%\hbox{
\includegraphics[width=16.0 cm, height=5.5 cm]{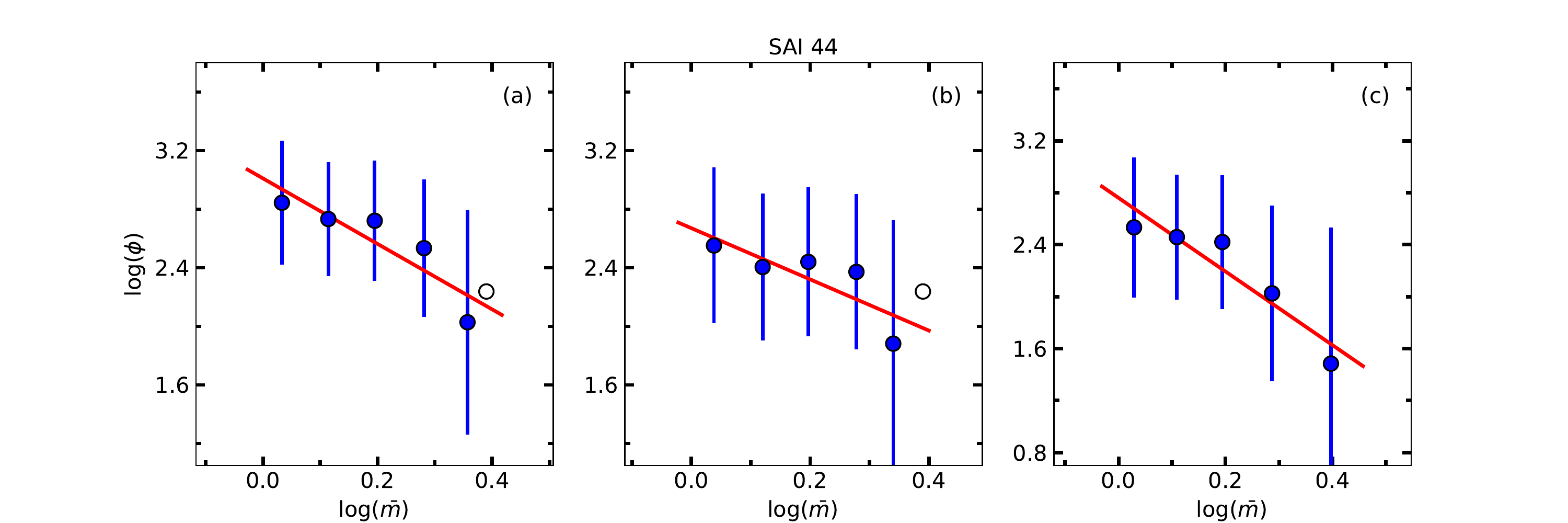}
%}
\caption{Plots of mass function slopes for SAI 44. Plots of entire regions, inner regions, and outer regions are given in (a), (b), and (c), respectively. The open circles in plots denote points excluded in the mass function slope determination of SAI 44.}
\label{mf}
\end{figure*}
%----------------------------------------------------

The number distribution of member stars of an open cluster according to the magnitude of the stars is called luminosity function (LF).  We calculated LF for the clusters SAI 44 and SAI 45 in V band down to 19 mag where our photometry is complete which is given in Table~\ref{lf}. The number distribution of stellar masses during the star formation events is known as initial mass function (IMF). The IMF determines the number of massive stars which were born in a star formation process and determine the fate of the star forming region. The mass function calculations for open clusters is retrieved from many complex situations like stellar birth rates and life times correction as stars associated with open clusters are coeval stars. The mass function (MF) of a cluster is defined as the number distribution of stars with stellar masses. We calculated present-day MF as N (log m) $\propto$ m$^{\Gamma}$. The ${\Gamma}$ is MF slope calculated as
\begin{center}
${\Gamma}$ = $\dfrac{d \ log \ N(log \ m)}{d \ log \ m}$
\end{center}
In the above relation N log(m) is number density on logarithmic mass scale. The MF obtained for two clusters are given in Table~\ref{lf}. The masses of individual member stars were found by fitting \citet{2017ApJ...835...77M} isochrones on the observed CMDs of SAI 44 and SAI 45 for their estimated age, apparent distance modulus, and reddening. The slopes were found by least square fitting and plots for the MF slopes are shown in Figure~\ref{mf}. The MF slopes of SAI 44 were calculated for entire region, inner region, and outer regions. The inner region was taken as circular region with radius of 4.3 arcmin around the center of the circle. The observed region beyond inner region was taken as outer region. The radius of the inner region circle was taken such as the number of stars should be approximately the same in inner and outer region. We found different values of the MF slopes in the three regions hinting towards variation in the MF slopes as a function of radial distance from the center. The slopes were found as -2.23$\pm$0.66, -1.75$\pm$0.72, and -2.83$\pm$0.61 in the entire, inner, and outer region respectively for the same mass range of 2.375-0.978 M$_{\odot}$. Here, we excluded one point that corresponds to only one stars in the mass-range 2.459-2.426 M$_{\odot}$ in calculation of MF slopes of SAI 44 as shown in Figure~\ref{mf}. The MF slope for the inner region is found to be flatter than the MF slope in outer region for SAI 44 which suggests presence of mass segregation \citep[e.g.,][]{2020MNRAS.494.4713M}.

We found two-step power law for the MF slopes in case of SAI 45 as can be noticed in Figure~\ref{mf}. The mass at which turnover of the MF slopes happens is called turnover mass, m$_{t}$. The stars having mass above m$_{t}$ are called high mass stars while remaining stars are called low mass stars. We found MF slopes for the entire observed region of SAI 45 to be -2.58$\pm$3.20 in the mass range 2.167-1.202 M$_{\odot}$. The MF slope for high mass stars of SAI 45 is steeper than the \citet{1955ApJ...121..161S} value of -1.35 for mass range 0.4 M$_{\odot}$ to 10.0 M$_{\odot}$. The steeper MF slope indicates towards presence of relatively less number of massive stars. The massive stars evolve faster than low mass stars so present day MF corresponds to a depletion of massive stars compared to IMF. The extent of massive stars depletion depends on initial condition of star formation \citep{2019MNRAS.489.2377S}. The turnover mass, m$_{t}$, for SAI 45 is found to be 1.26 M$_{\odot}$. The m$_{t}$ value for SAI 45 is higher than the value $\sim$1.06 M$_{\odot}$ found for NGC 381 in our previous study \citep{2020MNRAS.494.4713M}. The two-step power law of MF slope having turnover around 1 M$_{\odot}$ seems to be commonly present in open clusters as reported in several past studies \citep[e.g.,][]{2013MNRAS.434.3236K,2019MNRAS.489.2377S}. We could not calculate MF slopes separately for inner and outer regions for SAI~45 a it has very few member stars (74) in entire observed region which further dividing into inner and outer regions could have lead to statistically insignificant results.
%-----------------------------------------------------------
\begin{figure}
\centering
\includegraphics[width=8 cm, height=7.0 cm]{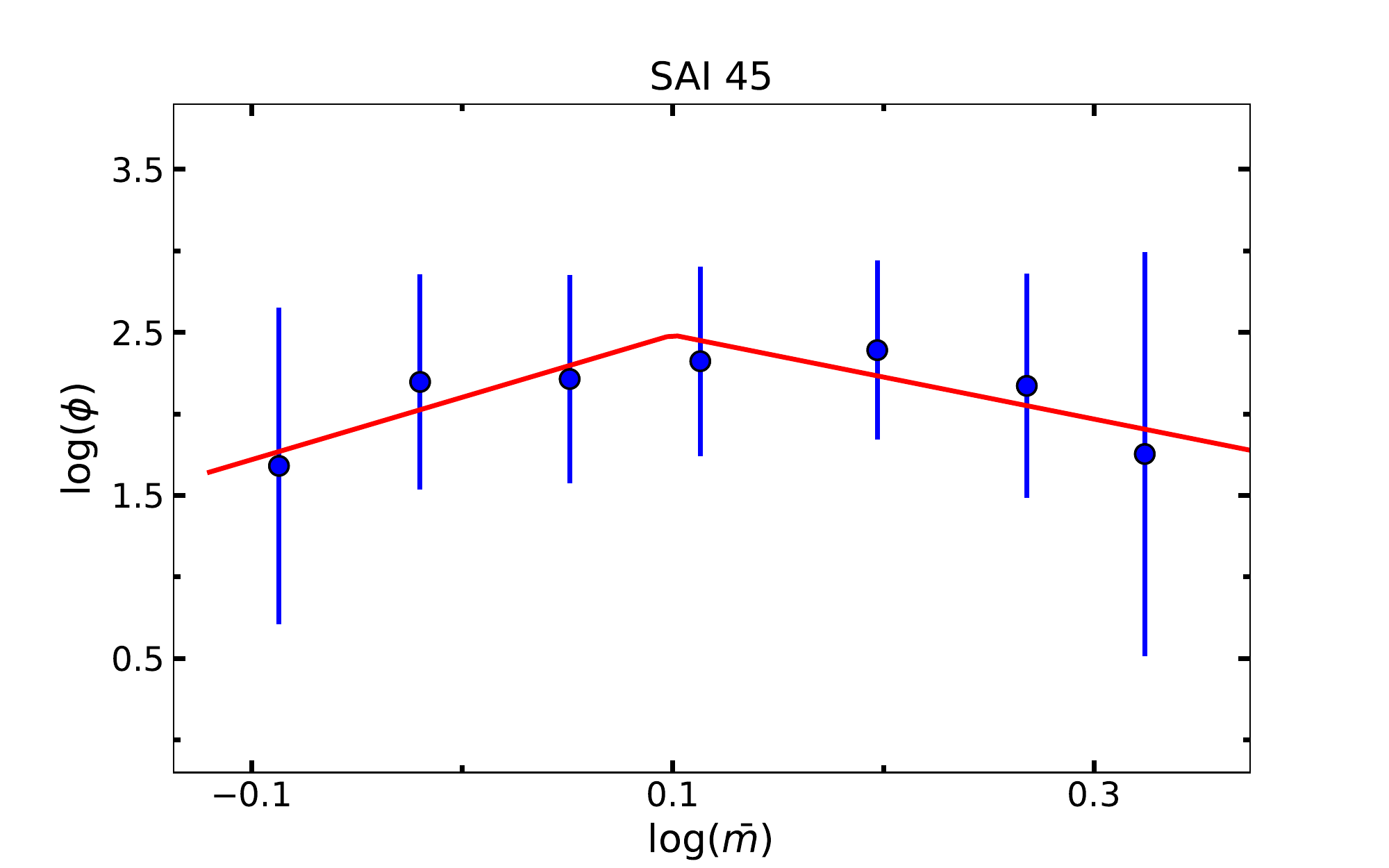}
\caption{MF for entire observed region of the cluster SAI 45.}
\label{mf_slope}
\end{figure}
%----------------------------------------------------

\begin{figure}
\includegraphics[width=9.0 cm, height=6 cm]{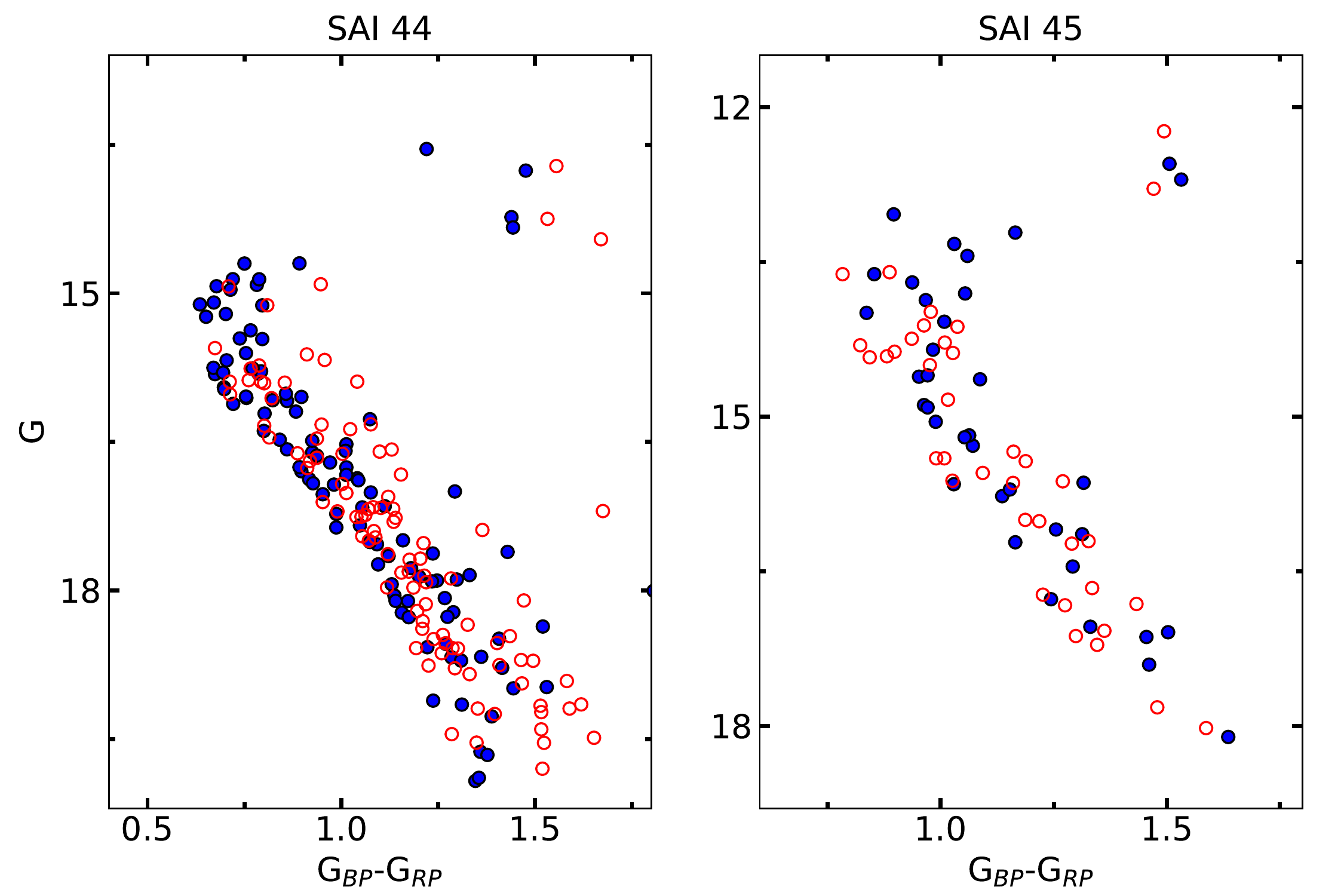}
\caption{The G/(G$_{BP}$-G$_{RP}$) CMD for visual inspection of mass segregation in SAI 44 and SAI 45. The member stars lying in inner region are shown by blue filled circular points whereas red open circles represent member stars belonging to outer region of the clusters.}
\label{m_seg}
\end{figure}

We noticed steeper MF slope in outer region of SAI 44 indicating mass segregation which can be attributed to the escape of the low mass stars from the cluster besides concentration of massive stars in the central region of the cluster caused by equi-partition of energy. The method based on the cumulative distribution of stars with radial distance for various mass bins may give misleading result due to dependence on the size of mass bins and cumulative radii.  We therefore used a method given by \citet{2009MNRAS.395.1449A} based on mass segregation ratio (MSR) to analyse the presence of mass segregation in the clusters. This method uses minimum sampling tree (MST) of the data points to determine the mean edge length $\gamma$. The MST of a sample of points is defined as the shortest path connecting all the points without any closed loops \citep{1957BSTJ...36.1389P}. We first calculated mean edge length for n most massive stars $\gamma_{mm}$ followed by mean edge length of n random stars of the whole sample $\gamma_{rand}$. We repeated the same calculation 500 times to get $\langle \gamma_{rand} \rangle$. Using the relation given by \citet{2011A&A...532A.119O}, we calculated mass segregation ratio $\Gamma_{MSR}$ as follows:
$$
\Gamma_{MSR} = \frac{\langle \gamma_{rand} \rangle}{\gamma_{mm}}
$$
The associated standard deviation $\Delta\Gamma_{MSR}$ with mass segregation ratio $\Gamma_{MSR}$ was calculated as:
$$
\Delta\Gamma_{MSR} = \Delta\gamma_{rand}
$$
The method basically uses assumption that any mass segregation in a cluster will be marked by relatively closer spatial distribution of massive stars than the low mass stars. $\Gamma_{MSR}$ $\sim$ 1 will mean that the spatial distribution of the most massive stars would be similar to the spatial distribution of low mass stars i.e. no mass segregation in the cluster. $\Gamma_{MSR}$ $>$ 1 would implies that the most massive stars are relatively closer than rest of stars which means presence of mass segregation. $\Gamma_{MSR}$ $<$ 1 indicates reverse mass segregation \citep{2018MNRAS.473..849D}. We found mass segregation ratio $\Gamma_{MSR}$ values as 1.22$\pm$0.17 and 1.07$\pm$0.23 for SAI 44 and SAI 45, respectively. The $\Gamma_{MSR}$ = 1.22$\pm$0.17 indicates presence of mass segregation in the clusters SAI 44. However, $\Gamma_{MSR}$ = 1.07$\pm$0.23 for SAI 45 suggests presence of weak or no mass segregation in the cluster. Being an old age cluster, SAI 45 is expected to have evidence of strong mass segregation. The dynamical time of these clusters is useful in understanding the reason behind the discrepancy of mass segregation levels in these clusters. SAI 44 was found to be dynamically older than SAI 45 as discussed in Section~\ref{tidal}, and hence, may have become more mass segregated \citep{2019ApJ...877...37R}. The tidal stripping of outer low mass stars may be another reason behind absence of mass segregation in SAI 45 \citep{2019ApJ...877...37R}. As number of member stars in SAI 45 are only 74 so our result regarding mass segregation is statistically preliminary in nature. We have shown G/(G$_{BP}$-G$_{RP}$) plots in Figure~\ref{m_seg} for visual inspection of mass segregation in the clusters. The radius of inner regions was taken as 4$^{\prime}$.3 and 4$^{\prime}$.5 for SAI 44 and SAI 45, respectively. The upper MS is dominated by stars belonging to inner region while lower MS is dominated by outer region stars in case of SAI 44. However, we could not find any conspicuous pre-eminence of stars belonging to either regions in the whole MS of SAI 45.  
%--------------------------------------
%-----------------

The dynamical relaxation time is a good indicator to find whether mass segregation is primordial or it happens because of dynamical relaxation \citep{2019MNRAS.488.1635A}. The dynamical relaxation time, T$_{E}$, is physically defined as the time interval in which velocity distribution of stars in a cluster approaches velocity distribution of Maxwellian equilibrium through exchange of energy among member stars. The dynamical relaxation time was calculated using \citet{1971ApJ...164..399S} formula as described in \citet{2020MNRAS.495.2496M}. We found T$_{E}$ values as 48 and 43 Myr for SAI 44 and SAI 45, respectively. We obtained T$_{E}$ values many times less compared to the obtained ages of the clusters which points towards dynamically relaxed state of SAI 44 and SAI 45. Thus, mass segregation in SAI 44 is possibly caused by dynamical relaxation.

%------------------------------------------------------------
\section{Kinematical structure}\label{kinematic}
%------------------------------------------------------------

We studied kinematics for the clusters using computational algorithm as described by \citet{2018Ap&SS.363...58E}. The line of sight velocities, V$_{r}$, are given by \citet{2020A&A...640A.127Z} as 2.20$\pm$5.73 and -33.28$\pm$35.30 km~s$^{-1}$ for SAI 44 and SAI 45, respectively. The value of V$_{r}$ for SAI 44 is found to be 7.33$\pm$0.05 km s$^{-1}$ by \citet{2019A&A...623A..80C} while \citet{2018A&A...619A.155S} reported V$_{r}$ values as 11.18$\pm$3.21 and -31.02$\pm$0.35 km s$^{-1}$ for SAI 44 and SAI 45, respectively. We used V$_{r}$ values as 7.33$\pm$0.05 and -31.02$\pm$0.35 km s$^{-1}$ in our present analysis from the above mentioned studies. The estimation of apex position and spatial velocities of the clusters are described below.

%----------------------------------------------------
\begin{figure}
\centering
\vspace{-0.3cm}
\includegraphics[width=8.5 cm, height=5 cm]{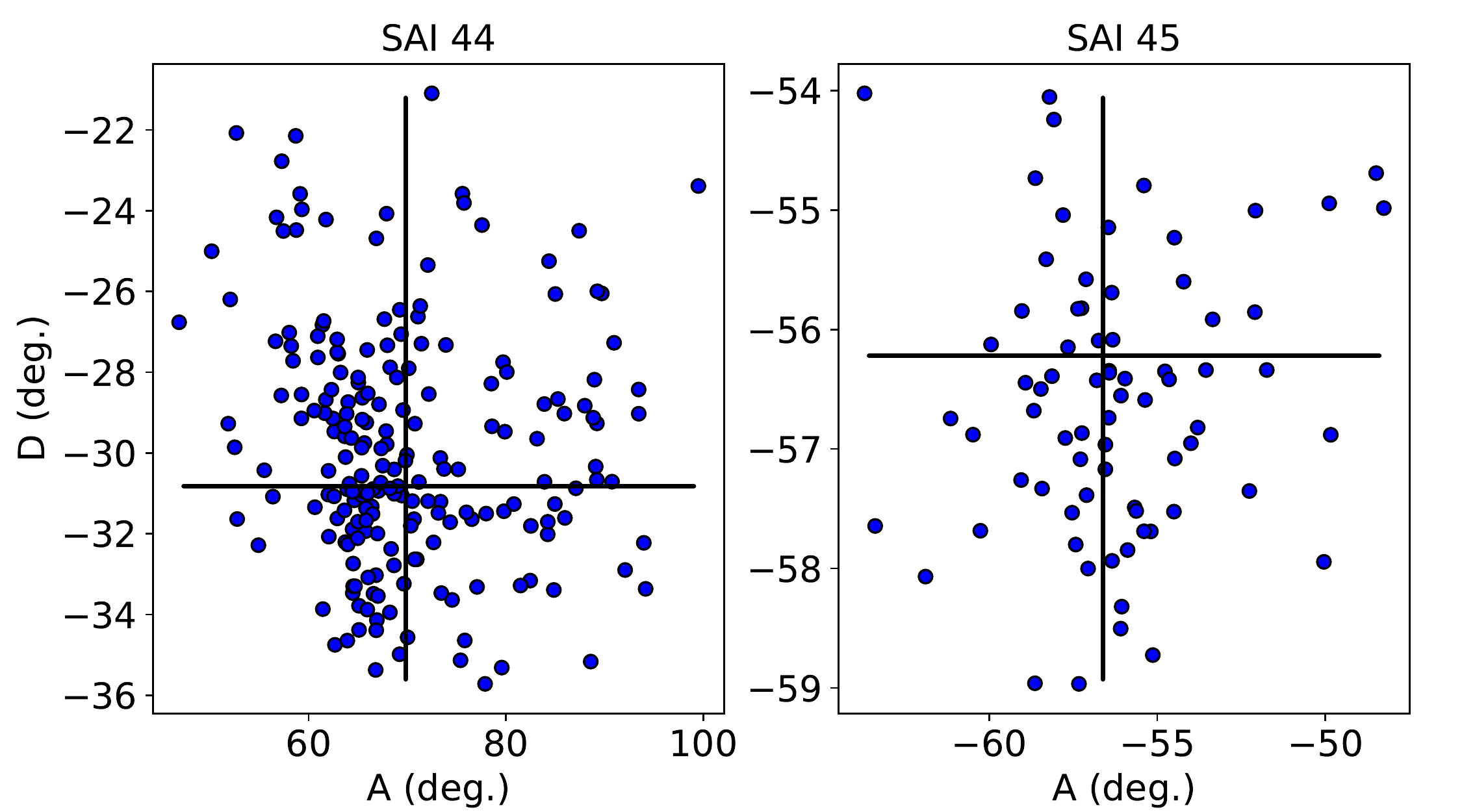}
\caption{Plots of AD-diagrams for SAI 44 and SAI 45.}
\label{AD}
\end{figure}
%-----------------------------------------------------
\begin{figure*}
\centering
\includegraphics[width=17 cm, height=5 cm]{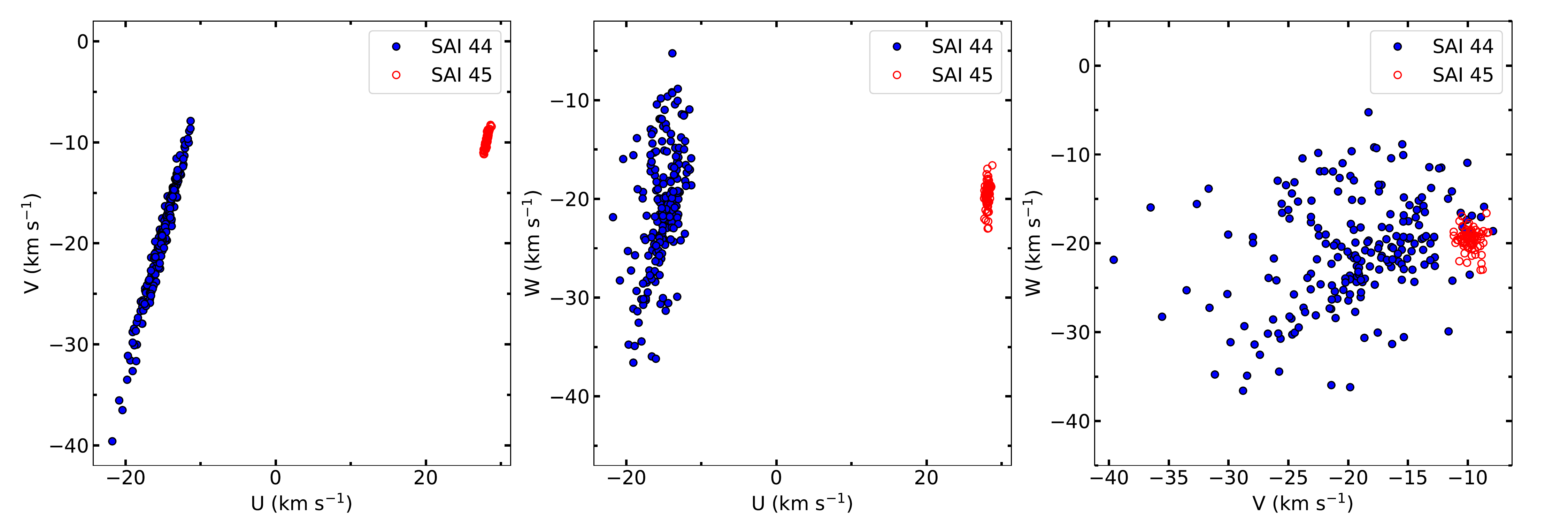}
\caption{Plots of space velocity components for SAI 44 and SAI 45. The blue circular points denotes space velocity components of SAI 44 member stars while space velocity components of SAI 45 member stars are shown by red open circles.}
\label{space_v}
\end{figure*}
%
%------------------------------------------------------------
\subsection{Apex of the clusters}
%------------------------------------------------------------
The apex position represents the motion of the clusters on the celestial sphere. As a open clusters is gravitationally bound system of stars therefore member stars of a cluster move with a common velocity vector and the parallel motions of the common velocity vector on the celestial sphere will point towards a coherent point known as the apex of the cluster. There are two methods to calculate apex position. The classical method is known as classical convergent point (CP) method while other is known as AD diagram method. These methods are based on the assumptions that a cluster is a non-rotating body without any expansion or contraction.\\

 i). The CP method:
  
  This method assumes that all the member stars of a cluster possess the same space velocities. The velocity components along the x, y, and z axes (V$_{x}$, V$_{y}$, V$_{z}$) for a cluster having coordinates ($\alpha$,$\delta$) with proper motions ($\mu_{\alpha*}$; $\mu_{\delta}$) and radial velocity V$_{r}$ at a distance of d$_{i}$ (in pc) in the heliocentric coordinate system are given by \citet{1968stki.book.....S} as described in \citet{2020RAA....20...16P}. We found the coordinates of the apex (A$_{CP}$, D$_{CP}$) to be (77$\fdg79\pm0\fdg11$, -45$\fdg71\pm0\fdg15$) for SAI 44 and (-78$\fdg55\pm0\fdg10$, -45$\fdg26\pm0\fdg10$) for SAI 45.\\
     
  ii). The AD diagram method:
  
  This method uses radial velocity and parallax of the stars. We can identify the group of stars having common space motions using this method. This method is described in good details by \citet{2006A&A...451..909C}. In the AD-diagram the positions of the apexes of the individual stars are calculated in equatorial coordinates. The (A, D) values of the individual member stars notify the positions of these stars through space velocity vectors. In this method intersection point (A$_{\circ}$, D$_{\circ}$) also called as apex in equatorial coordinates can be given as:
$$A_{\circ}=\tan^{-1}\Big[\frac{\overline{V_y}}{\overline{V_x}}\Big]$$
$$D_{\circ}=\tan^{-1}\Big[\frac{\overline{V_z}}{\sqrt{\overline{V_x}^2+\overline{V_y}^2}}\Big]$$
Coordinates of the apex (A$_{\circ}$, D$_{\circ}$) through AD-diagram method were calculated as (69$\fdg79\pm0\fdg11$, -30$\fdg82\pm0\fdg15$) and (-56$\fdg22\pm0\fdg13$, -56$\fdg62\pm0\fdg13$) for SAI 44 and SAI 45, respectively. The AD-diagrams for the two clusters are shown in Figure~\ref{AD}.
%
%--------------------------------------------------------------------------------
\begin{table*}
\caption{Kinematical parameters of SAI 44 and SAI 45 open clusters. References: 1- present study; 2- \citet{2018A&A...619A.155S}; 3- \citet{2018A&A...618A..93C}}
\centering
\hbox{
\begin{tabular}{lccc}
\hline
Parameters & SAI 44 & SAI 45 & References \\
&(FSR 716) & (FSR 727)&\\
\hline
\noalign{\smallskip}
No. of members (N)	&	204	&	74	&	1	\\
$\mu_{\alpha*}$ (mas/yr)	&	-0.20$\pm$0.25	&	-1.68$\pm$0.08	&	1	\\
$\mu_{\delta}$ (mas/yr)	&	-1.63$\pm$0.22	&	-1.33$\pm$0.08	&	1	\\
Distance d(pc)  & 3670$\pm$184 & 1668$\pm$47& 1\\
V$_{\alpha}$ (km/s) & -3.85 & -13.78 & 1\\
V$_{\delta}$ (km/s) & -31.19 & -10.87 & 1\\
V$_{t}$ (km/s) & 31.80 & 17.57 & 1\\
V (km/s) & 32.27 & 35.64 & 1\\
(A, D)$_{CP}$&     77$^{o}$.79$\pm$0$^{o}$.11, -45$^{o}$.71$\pm$0$^{o}$.15 & -78$^{o}$.55$\pm$0$^{o}$.10, -45$^{o}$.26$\pm$0$^{o}$.10& 1 \\
(A$_{o}$, D$_{o}$)&     69$^{o}$.79$\pm$0$^{o}$.11, -30$^{o}$.82$\pm$0$^{o}$.15 & -56$^{o}$22$\pm$0$^{o}$.13, -56$^{o}$.62$\pm$0$^{o}$.13 & 1 \\
($\bar{V}_{x}$, $\bar{V}_{y}$, $\bar{V}_{z}$)\hspace{0.1 cm}	\textrm{(km/s)}&	9.57$\pm$3.09, 26.01$\pm$5.10, -16.53$\pm$4.07	&	10.90$\pm$3.30, -16.30$\pm$4.04, -29.76$\pm$5.46 &	1	\\
($\bar{U}$, $\bar{V}$, $\bar{W}$)\hspace{0.1 cm} \textrm{(km/s)}	&	-15.14$\pm$3.90, -19.43$\pm$4.41, -20.85$\pm$4.57	&	28.13$\pm$5.30, -9.78$\pm$3.13, -19.59$\pm$4.43	&	1	\\
	&	-17.78$\pm$3.06, -14.85$\pm$1.13, -19.08$\pm$0.64	&	28.09$\pm$0.34, -9.68$\pm$0.14, -18.80$\pm$0.22	&	2\\
(x$_{c}$, y$_{c}$, z$_{c}$) (pc)	&	597, 2758, 2893 & 229, 1188, 1234	&	1	\\
X$_{\odot}$ (kpc)	&	-3.489$\pm$0.059	&	-1.590$\pm$0.039 	&	1	\\
	&	-3.386	&	-1.614	&	3	\\
	
Y$_{\odot}$ (kpc)	&	1.116$\pm$0.033 &	0.489$\pm$0.022	&	1	\\
	&	1.084	&	0.496	&	3	\\
	
Z$_{\odot}$ (kpc)	&	0.231$\pm$0.015	&	0.125$\pm$0.011	&	1	\\
	&	0.225	&	0.127	&	3	\\
R$_{gc}$ (kpc)	&	11.744$\pm$0.108	&	9.803$\pm$0.099	&	1	\\
	&	11.776	&	9.967	&	3	\\
\hline
\label{kinematic_sai}
\end{tabular}
}
\end{table*}

%-------------------------------------------------------------------
%
\subsection{Space velocity of the clusters}
%------------------------------------------------------------
We calculated space velocity components of the clusters SAI 44 and SAI 45 using space velocity components (V$_{x}$, V$_{y}$, V$_{z}$) along x, y, z-axes of the heliocentric coordinate system. We used \citet{2011A&A...536A.102L} relations to calculate space velocity components (U, V, W). The \citet{2011A&A...536A.102L} relations are described as following:
$$
U=-0.0518807421 V_{x}-0.8722226427 V_{y}-0.4863497200 V_{z}
$$
$$
V=+0.4846922369 V_{x}-0.4477920852 V_{y}+0.7513692061 V_{z}
$$
$$
W=-0.8731447899 V_{x}-0.1967483417 V_{y}+0.4459913295 V_{z}
$$
We calculated mean space velocity components ($\bar{U}$, $\bar{V}$, $\bar{W}$) form the values of space velocity components (U, V, W) obtained using above relations. The mean values ($\bar{U}$, $\bar{V}$, $\bar{W}$) were obtained as (-15.14$\pm$3.90, -19.43$\pm$4.41, -20.85$\pm$4.57) and (28.13$\pm$5.30, -9.78$\pm$3.13, -19.59$\pm$4.43) in km s$^{-1}$ for SAI 44 and SAI 45, respectively. The distribution of these space velocities are shown in Figure~\ref{space_v}. It can be inferred from the figure that SAI 44 has more extended velocities compared to a very compact velocity distribution of SAI 45 in (UVW) velocity space consistent with our finding that SAI 45 is a compact cluster in Section~\ref{tidal}.

%------------------------------------------------------------
\subsection{Other kinematic structure parameters}
%------------------------------------------------------------
The coordinate of the center ($x_{c}, y_{c}, z_{c}$) of the cluster is calculated through finding the center of mass of the N of member stars in the equatorial coordinates. The relations to calculate ($x_{c}, y_{c}, z_{c}$) are given below:
$$x_c=\left[\sum\limits_{i=1}^{N}d_i\cos\alpha_i\cos\delta_i\right]\Bigg{/}N$$
$$y_c=\left[\sum\limits_{i=1}^{N}d_i\sin\alpha_i\cos\delta_i\right]\Bigg{/}N$$
$$z_c=\left[\sum\limits_{i=1}^{N}d_i\sin\delta_i\right]\Bigg{/}N$$
We obtained ($x_{c}, y_{c}, z_{c}$) to be (597, 2758, 2893) and (229, 1188, 1234) in parsec for SAI 44 and SAI 45, respectively.

The Galactocentric distance of a cluster is calculated as using relation R$_{gc}^{2}$=R$_{\odot}^{2}$+(d$\cos{b}$)$^2$-2R$_{\odot}$d$\cos{b}\cos{l}$ where distance of the Sun from the Galactic center R$_{\odot}$ was taken as 8.2$\pm$0.10 kpc. \citep{2019MNRAS.486.1167B}. We obtained R$_{gc}$ values as 11.744$\pm$0.108 and 9.803$\pm$0.099 kpc for SAI 44 and SAI 45, respectively. The projected distances (X$_{\odot}$, Y$_{\odot}$, Z$_{\odot}$) in the Galactic plane can be calculated as follows:
$$
X_{\odot}=d\cos{b}\cos{l};
~~~Y_{\odot}=d\cos{b}\sin{l};
~~~Z_{\odot}=d\sin{b}
$$
where d is heliocentric distance while l and b are the Galactic longitude and latitude. We found (X$_{\odot}$, Y$_{\odot}$, Z$_{\odot}$) for SAI 44 and SAI 45 to be (-3.489$\pm$0.059, 1.116$\pm$0.033, 0.231$\pm$0.015) and (-1.590$\pm$0.039, 0.489$\pm$0.022, 0.125$\pm$0.011) in kpc, respectively. The values of all the kinematic parameters calculated by us are tabulated in Table~\ref{kinematic_sai}. 

%%%%%%%%%%%%%%

\section{Conclusion}\label{conclusion}
We presented the first deep BVR$_{c}$I$_{c}$ photometric and kinematic study of the open clusters SAI 44 and SAI 45. The study is based on the member stars identified taking advantage of kinematic data of \textit{Gaia} eDR3. We supplemented our photometric data with \textit{Gaia} eDR3 and Pan-STARRS PS1 DR2. The physical parameters like distance, total-to-selective extinction, reddening, age, half mass radius, and tidal radius are estimated. The total-to-selective extinction R$_{v}$ values are found to be following correlation between R$_{v}$ and distance as R$_{v}$ is smaller for cluster SAI 45 which is relatively at smaller distance. The ratio of half mass radius to tidal radius indicates that SAI 45 is a compact cluster which was found to be true from structural and kinematic study. It is found that the structure and dynamics of these clusters are influenced by tidal interaction. The main sequence of SAI 45 hosts eMSTO which corresponds to an apparent age spread of 493 Myr. We calculated possible apparent spread in age caused by photometric uncertainty and metallicity spread and found that any possible apparent age spread due to these two factors would be very smaller than the actual apparent age spread inferred from the eMSTO. We found i) apparent age spread similar to the  apparent age spread and cluster age relation predicted by rotation models, ii) SYCLIST synthetic population with different rotation rates was able to reproduce observed eMSTO, iii) stars in red part of eMSTO were preferentially concentrated in inner region. These findings support the theory attributing origin of eMSTO to the differential rotation of eMSTO stars. Therefore we argue that the presence of eMSTO in SAI 45 is a stellar evolution rather than star formation phenomenon mainly caused by differential rotation rates of stars. We obtained steeper MF slope in outer region for SAI 44 indicating presence of mass segregation in the cluster. We also studied these clusters for mass segregation using minimum sampling tree method and found that mass segregation is present in SAI 44. However, we could not get any evidence of strong mass segregation in SAI 45 possibly due tidal striping of outer low mass stars. We calculated kinematic structure parameters of these clusters and the obtained parameters were used in understanding the dynamical evolution of these clusters. The main results of the present study are given below:
\begin{enumerate}
\item We calculated spatial stellar density distribution and found that SAI 44 and SAI 45 are sparse clusters with irregular shape using density contrast parameter and contour map of spatial stellar density.
\item We calculated membership probabilities of stars based on statistical method using proper motions. We found 204 and 74 member stars in SAI 44 and SAI 45, respectively. We used member stars exclusively to determine physical and dynamical parameters of these clusters.
\item The ages were determined through isochrone fitting on the color-magnitude diagrams. We obtained log(age) to be 8.82$\pm$0.10 and 9.07$\pm$0.10 years for SAI 44 and SAI 45, respectively.
\item We studied extended main-sequence turn-off present in color-magnitude diagram of SAI 45 which is mainly caused by different rotation rates of member stars.
\item We found relatively steeper mass function slopes equal to -1.75$\pm$0.72 and -2.58$\pm$3.20 for stars in the entire observed region of SAI 44 and SAI 45, respectively. We also found steeper mass function slope in outer region than the same in inner region of SAI 44 which can be signature of the presence of mass segregation in the cluster.
\item We found that the dynamical relaxation times for SAI 44 and SAI 45 are many times less than the age of these clusters. Thus, these clusters have achieved dynamical relaxation through dynamical evolution.
\item Assuming that the clusters are non-rotating bodies without any contraction or expansion we calculated kinematic parameters using radial velocities of the clusters. We found apex position of SAI 44 and SAI 45 using AD-diagram to be (69$\fdg79\pm0\fdg11$, -30$\fdg82\pm0\fdg15$) and (-56$\fdg22\pm0\fdg13$, -56$\fdg62\pm0\fdg13$), respectively. We obtained space velocities components in km s$^{-1}$ for SAI 44 and SAI 45 to be (-15.14$\pm$3.90, -19.43$\pm$4.41, -20.85$\pm$4.57) and (28.13$\pm$5.30, -9.78$\pm$3.13, -19.59$\pm$4.43), respectively.
\end{enumerate} 

\section{ACKNOWLEDGEMENTS}
Data from Pan-STARRS surveys (PS1) was used in this paper. The Pan-STARRS1 Surveys (PS1) and the PS1 public science archive have been made possible through contributions by the Institute for Astronomy, the University of Hawaii, the Pan-STARRS Project Office, the Max-Planck Society and its participating institutes, the Max Planck Institute for Astronomy, Heidelberg and the Max Planck Institute for Extraterrestrial Physics, Garching, The Johns Hopkins University, Durham University, the University of Edinburgh, the Queen's University Belfast, the Harvard-Smithsonian Center for Astrophysics, the Las Cumbres Observatory Global Telescope Network Incorporated, the National Central University of Taiwan, the Space Telescope Science Institute, the National Aeronautics and Space Administration under Grant No. NNX08AR22G issued through the Planetary Science Division of the NASA Science Mission Directorate, the National Science Foundation Grant No. AST-1238877, the University of Maryland, Eotvos Lorand University (ELTE), the Los Alamos National Laboratory, and the Gordon and Betty Moore Foundation. This work presents results from the European Space Agency (ESA) space mission Gaia. Gaia data are being processed by the Gaia Data Processing and Analysis Consortium (DPAC). Funding for the DPAC is provided by national institutions, in particular the institutions participating in the Gaia MultiLateral Agreement (MLA).

%-------------------------------------------------------------------
\bibliographystyle{yahapj}
\bibliography{main}

\end{document}